\documentclass[journal]{new-aiaa}
\usepackage[utf8]{inputenc}

\usepackage{graphicx}
\usepackage{amsmath}
\usepackage[version=4]{mhchem}
\usepackage{siunitx}
\usepackage{longtable,tabularx}
\usepackage{multirow}
\usepackage{multicol}
\usepackage{algorithm}
\usepackage{algpseudocode}
\usepackage{changepage}
\title{Aerodynamic Shape Design Space Exploration with Deep Latent Diffusion Model}

\author{Zhen Wei~\footnote{Doctoral Student, Computer Vision Lab, School of Computer and Communication Sciences, BC Building, Station 14, Lausanne, Vaud, 1015, Switzerland. Student Member AIAA. }}
\author{Edouard Dufour~\footnote{Doctoral Student, Computer Vision Lab, School of Computer and Communication Sciences, BC Building, Station 14, Lausanne, Vaud, 1015, Switzerland.}}
\author{Colin Pelletier~\footnote{Master Student in Communication Systems, School of Computer and Communication Sciences, BC Building, Station 14, Lausanne, Vaud, 1015, Switzerland.}}
\affil{Swiss Federal Institute of Technology in Lausanne (EPFL), Lausanne, Vaud, 1015, Switzerland}
\author{Michaël Bauerheim~\footnote{Associate Professor, Department of Aerodynamics, Energies and Propulsion, 10 Avenue Marc Pélegrin, Toulouse, Occitania, 31055, France.}}
\affil{Higher Institute for Aeronautics and Space (ISAE-SUPAERO), Toulouse, Occitania, 31055, France}
\author{Pascal Fua~\footnote{Professor, Computer Vision Lab, School of Computer and Communication Sciences, BC Building, Station 14, Lausanne, Vaud, 1015, Switzerland; pascal.fua@epfl.ch. (Corresponding Author)}}
\affil{Swiss Federal Institute of Technology in Lausanne (EPFL), Lausanne, Vaud, 1015, Switzerland}

\newif\ifdraft
\draftfalse
\drafttrue

\usepackage[dvipsnames]{xcolor}
\definecolor{orange}{rgb}{1,0.5,0}
\definecolor{violet}{RGB}{70,0,170}

\ifdraft
 \newcommand{\PF}[1]{{\color{blue}{\bf PF: #1}}}
 
 \newcommand{\WZ}[1]{{\color{SeaGreen}{\bf ZW: #1}}}
 \newcommand{\wz}[1]{{\color{SeaGreen} #1}}
 \newcommand{\MB}[1]{{\color{blue}{\bf MB: #1}}}
 
 \newcommand{\BG}[1]{{\color{orange}{\bf BG: #1}}}
 
 \usepackage{ulem}
\else
 \newcommand{\PF}[1]{}
 
 \newcommand{\WZ}[1]{}
 \newcommand{\wz[1]{ #1 }
 \newcommand{\MB}[1]{}
 
 \newcommand{\BG}[1]{}
 
\fi

\newcommand{\comment}[1]{}

\newcommand{\acron}[0]{\textit{DiffGeo}}

\newcommand{\bI}{\mathbf{I}}

\newcommand{\bv}{\mathbf{v}}

\newcommand{\bw}{\mathbf{w}}

\newcommand{\bz}{\mathbf{z}}

\newcommand{\bmu}{\mathbf{\mu}}

\newcommand{\bZ}{\mathbf{Z}}

\newcommand{\cL}{\mathcal{L}}
\newcommand{\cF}{\mathcal{F}}
\newcommand{\cN}{\mathcal{N}}
\newcommand{\cD}{\mathcal{D}}

\newcommand{\cT}{\mathcal{T}}

\newcommand{\argmin}{\operatornamewithlimits{argmin}}

\newcommand{\revise}[1]{#1}
\newcommand{\revisetwo}[1]{#1}

\begin{document}

\vspace*{-2.7cm}
\begin{adjustwidth}{-1.0cm}{-1.0cm}
    \noindent
    {\sffamily\bfseries\large POSTPRINT NOTICE} \\
    \noindent
    {\sffamily\fontsize{8.5}{8}\selectfont
        This is the authors' postprint, not the Version of Record, of the \textit{AIAA Journal} article published online on 31 Aug. 2026
    (\href{https://doi.org/10.2514/1.J066320}{10.2514/1.J066320}).
        \par
        \noindent
        \textbf{Data, code and agentic skill demo:}
        \href{https://github.com/kfxw/DiffGeo}{https://github.com/kfxw/DiffGeo}
        \par
    }
    \noindent
    {\sffamily\fontsize{8.5}{8}\selectfont
        Postprint posted on 31 August 2026.\par
        \noindent
        This document is under the Creative Commons Attribution 4.0 International
        (CC BY 4.0) license in accordance with the SNSF open-access requirement.
        \par
    }
\end{adjustwidth}
\vspace{-5mm}

\maketitle

\footnotetext{This work is an extension of the AIAA's conference paper: Zhen Wei, Edouard Dufour, Colin Pelletier, Pascal Fua and Michaël Bauerheim. "DiffAirfoil: An Efficient Novel Airfoil Sampler Based on Latent Space Diffusion Model for Aerodynamic Shape Optimization," AIAA 2024-3755. AIAA AVIATION FORUM AND ASCEND 2024, Las Vegas, NV, July 29-August 2, 2024.}

\vspace{-8mm}

\begin{abstract}
We propose \textit{DiffGeo}, a latent space diffusion-based generative framework for aerodynamic design space exploration under extreme data scarcity. \textit{DiffGeo} combines a learned latent space model for automatic shape parameterization, with a diffusion sampler to directly generate novel, geometry-valid and controllable designs. We validate the approach on a series of case studies: (i) a 2D airfoil generation benchmark, where \textit{DiffGeo}’s latent diffusion model is compared against GAN- and VAE-based baselines in terms of sample quality, diversity and constraint adherence under limited data; (ii) integration into a surrogate-based optimization pipeline, where \textit{DiffGeo}'s conditional sampling produces task-informed airfoil data that improve both surrogate modeling and optimization performance; and (iii) extension to 3D turbomachinery blade prototyping, where \textit{DiffGeo} generates realistic and high-performance blade geometries from a small set of reference designs. Throughout these investigations, \textit{DiffGeo} achieves high-quality and diverse shape generation with at least an order of magnitude less data than alternatives, decouples geometry representation from design targets for flexible reuse, and seamlessly incorporates complex design constraints via energy-based conditioning. These capabilities demonstrate \textit{DiffGeo}’s potential to enhance early-stage design by automating design space exploration--improving efficiency, expanding design diversity and embedding engineering knowledge through controllable guidance.

\end{abstract}

\section*{Nomenclature}
{\renewcommand\arraystretch{1.0}
\noindent\begin{longtable*}{@{}l @{\quad=\quad} l@{}}

$\mathcal{A}(V)$           & Airfoil area given ordered contour vertices $V$. \\
$AoA$           & Angle of attacks. \\
$\beta_1(r),\beta_2(r)$    & Spanwise blade metal angle distribution (twist law) at leading and trailing edges from hub to tip. \\
$C$                        & Scalar condition input of conditional GAN. \\
$C_L, C_D, C_P$                 & Lift, drag and pressure coefficients. \\
$C^E, C^I$                 & Equality and inequality constraints used in energy-based guidance function. \\
$\cD_{KL}$                 & Kullback–Leibler divergence. \\
$D^k_{intra},D^k_{inter}$  & $k$-nearest-intra/inter-sample distance to measure diversity and novelty. \\
$\cF$                      & Fréchet inception distance. \\
$f_\Theta$                 & Auto-decoder function of LSM with parameters $\Theta$. \\
$g_\Gamma^{(G)}, g_\Pi^{(D)}$ & Generator (parameters $\Gamma$) and discriminator (parameters $\Pi$) networks of the GAN baseline. \\
$g_\Phi$                   & LSDM with parameter $\Phi$. \\
$\cL$                      & Loss function for model training. \\
L/D                        & Lift-to-drag ratio. \\
$M = {V, E}$               & A shape represented as a mesh, with vertices $V$ and edges $E$. \\
$\hat{M} = {\hat{V}, E}$   & Template mesh used by LSM to deform from. \\
$q, p$                     & Forward and reverse process of denoising diffusion model. \\
$S$                        & Shape in training dataset. \\
$T$                        & Total number of diffusion timesteps. \\
$\cT(x)$                   & Airfoil thickness distribution at chordwise location $x$. \\
$Z,\bz$                    & Latent space and latent code learned by LSM. \\
$\bz_t$                    & Latent code at diffusion timestep $t$. \\
$\eta$                     & Isentropic efficiency of blade.

\end{longtable*}}

{\renewcommand\arraystretch{1.0}
\noindent\begin{longtable*}{@{}l @{\quad=\quad} l@{}}

\end{longtable*}}


\begin{figure}[tbh]
    \begin{center}
        \includegraphics[width=1\linewidth]{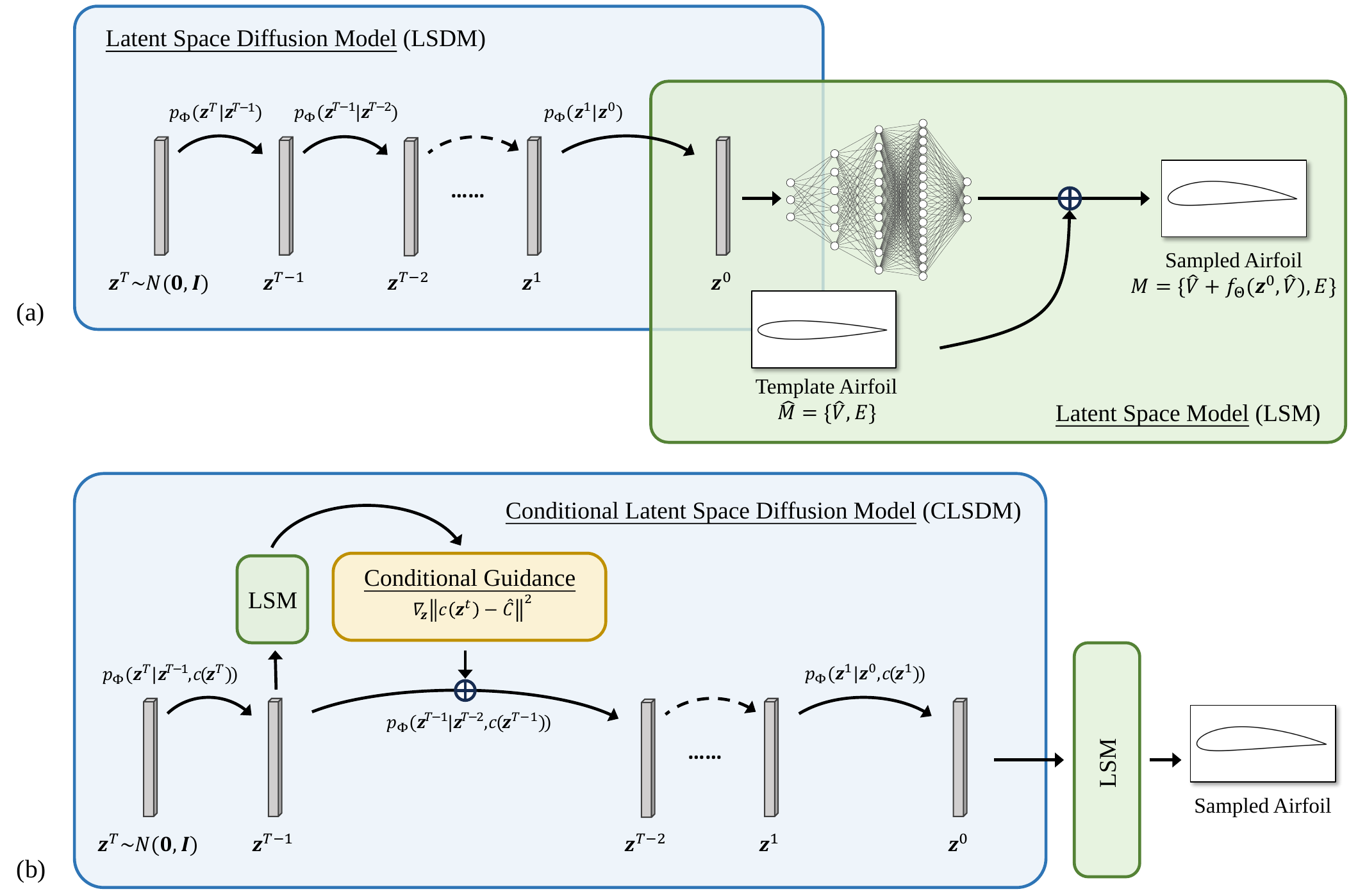}
    \end{center}
    \vspace{-4mm}
    \caption{
        \small \textit{DiffGeo} framework. (a) Unconditional \acron{} and (b) conditional \acron{}.
    }
    \label{fig:abs_framework}
\end{figure}

\section{Introduction}

Design Space Exploration (DSE) is the systematic process of sampling, evaluating, and selecting candidate designs across a high-dimensional geometry space to satisfy multiple objectives and constraints. In aerospace engineering, DSE constitutes a foundational step in aerodynamic shape optimization (ASO) and multidisciplinary design optimization (MDO) with many practical applications. Traditionally, DSE is implemented by combining low-dimensional, handcrafted parameterizations--such as Bézier curves, B-splines or Class/Shape Transformation (CST)~\cite{Kulfan08}--with statistical sampling techniques like Latin hypercube sampling (LHS)~\cite{McKay79} and surrogate modeling to approximate expensive performance evaluations. While effective, this paradigm can prove inefficient, inflexible and iterative for several reasons. First, traditional DSE workflows often requires dedicated shape modeling to ensure the generation of valid and related designs, including the choice of an appropriate shape parameterization and subsequent hyperparameter tuning. This process involves imposing extensive constraints on shape variables and managing the trade-off between design space exploration and exploitation. Second, these methods are usually restricted to low-dimensional design spaces, which limits their ability to capture complex geometric variations. Third, the usual reliance on Monte Carlo–style `sampling-evaluation-selection' often requires many random samples and performance evaluations to meet the design objectives, which is inefficient. This is because the sampling process is passive and cannot be actively guided toward feasible or high-performing regions of the design space. This lack of sampling control is particularly limiting when addressing complex, high-dimensional design objectives and constraints. This ultimately restricts the adaptability and effectiveness of the design process.

Recent advances in machine learning offer a promising way to overcome these limitations. Deep generative design methods train models to directly synthesize new designs by learning from prior examples through supervised, data-driven optimization. By capturing the joint distribution of feasible geometries, performance and constraints, these models can generate candidate designs that satisfy desired criteria in one shot. Early applications of deep generative models to aerodynamic design have demonstrated this potential. For example, conditional generative adversarial networks (GANs)~\cite{Goodfellow14} and variational autoencoders (VAEs)~\cite{Kingma14b} have been used to generate 2D airfoils conditioned on target aerodynamic properties~\cite{Chen20d,Li20i,Li21e,Du20,Achour20,Wang22c,Lei21,Yonekura21,Kou23,Swannet24}. These studies have confirmed the feasibility of learning-based shape generation but also suffer from some limitations.  First, most models require large training datasets, often on the order of $10^3$ samples for 2D airfoils for example, to generate meaningful and diverse designs. Such data demands can rarely be met in aerospace engineering, where diverse and representative shapes are costly to obtain. Second, existing generative models entangle geometry with performance in a way that limits adaptability, particularly given the highly customized nature of design requirements. In other words, a model trained for a specific task, such as low-speed, lift-conditioned airfoil generation, cannot be easily repurposed for different scenarios or objectives without collecting new data and retraining. Third, controllability over the generated designs remains limited. Most models only accept a few scalar values, such as  body forces or operating conditions, as conditioning inputs, making it difficult to impose more complex constraints, such as spanwise thickness distributions or 3D geometry profiling. In summary, current data-driven generative DSE approaches require a significant amount of data, cannot adapt easily to new design objectives, and do not provide fine-grained and effective controllability for design space exploration.

To address these challenges, we introduce \acron{}, a novel deep generative method for DSE in multidisciplinary design optimization. An early version of this approach was presented in a conference paper~\cite{Wei24b}. It focused on 2D airfoils and this paper generalizes it to a wider range of geometries and shows it can be turned into a widely applicable DSE tool. As shown in Fig.~\ref{fig:abs_framework}, \acron{} is built upon a diffusion-based generative model~\cite{SohlDickstein15} that operates on a shape latent space learned using an automated parameterization method~\cite{Wei23a,Wei23b}. When guided by task-specific energy functions, \acron{} can efficiently generate aerodynamic geometries satisfying complex design objectives.  We show that diffusion-based sampling offers more stable training than  GAN- and VAE-based approaches,  eliminating the need for adversarial discriminators and enabling data-efficient model training. Crucially, \acron{} is designed to provide a task-agnostic shape sampler independent of any performance target. When applied to new tasks, it allows rapid adaptation by simply replacing or combining one or more differentiable energy functions as guidance, without having to retrain the whole model. Notably, we introduce an enhanced conditional generation scheme to further improve the controllability of complex guidance.

In short, we aim to establish \acron{} as a novel generative framework for design space exploration. Our key contributions include:
\begin{itemize}
    \item \textbf{High data efficiency.} \textit{DiffGeo} can be effectively trained in data-scarce environments to construct a meaningful design space and to enable its exploration. For example, our experiments show successful training with as few as 50 airfoils or 75 3D blades linearly interpolated from six base profiles, which is one to two orders of magnitude less data than prior deep generative models. This reduces the data requirements by one or two orders of magnitude, compared  to GAN- and VAE-based approaches. By drastically lowering the data barrier, \textit{DiffGeo} enables generative DSE when large-scale geometry databases cannot be obtained, which is often the case in aerospace engineering.

    \item \textbf{Geometry-performance disentanglement.} \textit{DiffGeo} decouples geometry generation from task-specific objectives, making the core shape generator general-purpose and reusable. When adapting to different tasks, new objectives and constraints are imposed via energy functions during sampling, without retraining the underlying generator.

    \item \textbf{Controllability Under High-Dimensional Constraints.} \textit{DiffGeo} supports fine-grained, multi-parameter conditioning giving designers precise control over the generated geometry with complex, vector-valued conditions, such as twisting distributions along a 3D rotor blade. This enables exploration of designs that satisfy spatially varying constraints, a capability absent in existing generative design tools.
\end{itemize}
Although we focus on geometric design objectives in this study, the framework could be extended to performance-based objectives and this will be the subject of future work. 

The remainder of this paper is structured as follows. Section~\ref{sec:related_work} discusses the context of related methods in design space exploration, including conventional design of experiments, data-driven model and generation-based approaches. Section~\ref{sec:method} describes \acron{}'s technical details. Section~\ref{sec:exp} evaluates \acron{} across three case studies: (i) a 2D airfoil generation benchmark that compares \acron{} with GAN- and VAE-based methods on sample quality, diversity, and controllability under extreme low-data conditions; (ii) task-informed data generation with \acron{} for improving surrogate-based optimization; and (iii) 3D turbomachinery blade prototyping with strict geometric constraints and limited reference designs. Finally, Section~\ref{sec:conclusion} concludes the paper, discusses \acron{}'s broader impact and proposes directions for future work.


\section{Related Work}
\label{sec:related_work}

As a novel generative framework for DSE, \textit{DiffGeo} lies at the intersection of geometry modeling, deep generative learning and multi-disciplinary design optimization. To contextualize its contributions, we review related work across three areas: Design space exploration in aerospace, with a focus on geometry parameterization; deep generative models in ASO; and diffusion models.

\subsection{Design Space Exploration in Aerospace}
Design space exploration in aerospace has gone through several major stages, each marked by advances in geometry modeling and sampling methodology. The foundation was laid in the 1930s with the NACA airfoil series, where systematic variation of simple shape parameters, such as camber and thickness, enabled engineers to explore predefined families of airfoils~\cite{Jacobs33}. These early parametric studies provided smooth, physically interpretable shapes but constrained exploration to narrow subspaces of the design space. As computational methods matured, formal Design of Experiments (DoE)~\cite{Fisher35} techniques such as Latin Hypercube Sampling (LHS)~\cite{McKay79} were introduced to support broader and more uniform coverage of parameter spaces. CFD-based optimization studies, like \citet{Hicks78}, demonstrated the need for more flexible parameterizations that could balance geometric expressivity against numerical stability. The emergence of gradient-based optimization and adjoint sensitivity analysis then allowed high-dimensional shape manipulation, but continued to rely on analytic parameter bases like the Hicks-Henne bump functions, \revise{the Class/Shape Transformation (CST) method~\cite{Kulfan08}, or deformation approaches such as Free-Form Deformation (FFD)~\cite{Sederberg86, Lamousin94, Kenway10}}. While these parameterizations kept the design space tractable, they still imposed structural assumptions on shape variation. In response, surrogate-assisted DSE was introduced: Statistical models, such as Kriging~\cite{Matheron64}, response surfaces~\cite{Box51} and neural networks~\cite{Rumelhart86a}, were built from structured DoE samples, enabling global search with significantly fewer high-fidelity evaluations~\cite{Sacks89, Barthelemy93}. These methods expanded exploration beyond local gradients but still required carefully crafted parameterizations and iterative optimization loops.

More recently, the field has shifted toward data-driven modeling of the geometry space for DSE. Latent space representations derived from principal component analysis~\cite{Robinson01,Poole15,Masters17,Li19d,Li21k}, autoencoders~\cite{Agostino18,Rios21,Li20i}, variational autoencoders~\cite{Yonekura21,Kou23,Swannet24,Wang22c} or autodecoders~\cite{Wei23a, Wei23b} have been used to capture nonlinear shape variation across existing databases of feasible designs. These learned manifolds restrict sampling to valid and meaningful geometries, addressing the challenge that most random shape perturbations yield abnormal results.

While these latent‐space methods improve geometric expressivity, so far they have been paired primarily with rule-based static sampling or surrogate‐based search strategies. The exploration and exploitation of such DSE methods heavily rely on the design of the sampling strategy and the quality of the surrogate models. Deep generative models advance this line of work by learning a generative sampler, enabling the direct synthesis of novel designs within the learned space.

\subsection{Deep Generative Design for Aerodynamic Design}

Deep generative design methods have become increasingly popular in ASO and MDO, enabling rapid synthesis of novel geometries with minimal manual parameter tuning through data-driven training. Among these, \citet{Chen20d} first applied GAN-based models~\cite{Goodfellow14} to 2D airfoil generation by embedding Bézier-curve parameterizations. Subsequent efforts extended these approaches using B-spline parameterizations~\cite{Du20} and mode decomposition~\cite{Li20i,Li21e}. Conditional GANs have also been used to accommodate constrained design objectives~\cite{Achour20,Wang22c,Lei21}. While GANs can generate smooth and diverse samples, their adversarial training paradigm often suffers from instability, including mode collapse, and demands substantial data as well as architectural tuning~\cite{Arjovsky2017}.

Invertible neural networks (INNs)~\cite{Dinh15,Dinh17a} offer a complementary approach by learning bijective mappings between geometry and performance spaces. For example, \citet{Glaws22} demonstrated that INNs trained on airfoil–performance pairs can invert target performance metrics into multiple valid geometries in a single forward pass. However, such models require carefully curated paired datasets and strict balancing of forward and inverse loss functions, which introduces additional data engineering complexity.

More recently, diffusion-based models~\cite{Ho20a} have emerged in aerodynamic design. Our previous work \textit{DiffAirfoil}~\cite{Wei24b} was the first to introduce diffusion models into airfoil generation. It demonstrated the model's superiority over GANs under data-scarce conditions. Several follow-up studies explored conditional diffusion models for airfoil synthesis~\cite{Graves24,Yonekura24,Wagenaar24,Yang24}. However, these methods often adopt a Classifier-Free Guidance (CFG) mechanism that models the joint distribution of geometry and performance, which requires large-scale labeled datasets. This makes their training costs comparable to GANs or INNs. Furthermore, the lack of explicit geometric parameterization leads to instability in generation, often necessitating heavy post-processing or smoothing tricks to produce valid designs~\cite{Graves24,Yonekura24}.

Despite these advances, deep generative models still face challenges in training stability, data efficiency and controllability. Building on \textit{DiffAirfoil}, \textit{DiffGeo} aims to address these limitations. Rather than being as a monolithic model for end-to-end generative design, we position \textit{DiffGeo} as a tool to automate and enhance existing MDO workflows that produce high-quality design candidates for downstream analysis and optimization through rapid prototyping.

\subsection{Denoising Diffusion Models}

Modern denoising diffusion models~\cite{SohlDickstein15,Ho20a,song2021scorebased} have emerged as a powerful and stable generative framework over the past decade. These models operate by gradually corrupting data with Gaussian noise and then training a neural network to learn the reverse denoising process. Originally formulated using discrete-time Markov chains, recent advances reinterpret this process as solving stochastic differential equations (SDEs), which can be further reformulated into equivalent deterministic formulations such as probability flow ordinary differential equations (ODE)~\cite{song2021scorebased,Karras22}, and continuous-time normalizing flows~\cite{Lipman23,Liu23,Albergo23}. \revise{In the domain of image synthesis, \citet{Rombach22} demonstrated that performing the diffusion process in a compressed latent space rather than the high-dimensional pixel space significantly improves computational efficiency and training stability. \textit{DiffGeo} adopts a similar philosophy by performing diffusion within the compact manifold learned by LSM.}

Conditional diffusion models introduce guidance mechanisms to control the generation process toward desired outputs. \citet{Dhariwal21} proposed classifier guidance, which incorporates gradients from a pre-trained external classifier to bias diffusion sampling toward specific class labels, effectively trading off sample diversity for fidelity. As an alternative, \citet{Ho22a} introduced Classifier-Free Guidance (CFG), which jointly trains conditional and unconditional score estimators and interpolates between them at sampling time. While CFG has become the foundation of recent large-scale diffusion models, it typically requires massive datasets and training budgets, making it impractical for data-scarce domains such as aerospace engineering. In contrast, \textit{DiffGeo} develops the energy-based guidance that decouples geometry generation from performance evaluation through differentiable energy functions, enabling task-specific guidance at sampling time while improving data efficiency and model reusability.

Diffusion methods have also been applied to 3D shape modeling, initially targeting denoising directly on point clouds~\cite{Yang19d,Mao22}. However, the data-driven nature of these approaches, combined with the lack of explicit shape parameterizations, often leads to irregular or invalid geometries—such as noisy point clusters, disconnected components, or self-intersecting surfaces. These issues are especially problematic in aerospace engineering, where high shape validity is critical and oscillatory or non-physical shapes must be avoided~\cite{Zahid24}. Point cloud-based diffusion models struggle to guarantee watertight, smooth geometries suitable for downstream simulation and analysis. \textit{DiffGeo} addresses this limitation by introducing an automatic shape parameterization stage that learns a latent space constrained to valid geometry manifolds. The diffusion process is then learned within this latent space, which enables the generation of controlled 3D shapes that are inherently regular and simulation-ready, thus overcoming a key limitation of earlier diffusion and point cloud-based 3D modeling methods.


\section{Methodology }
\label{sec:method}

\textit{DiffGeo} is designed to sample novel designs using a shape parameterization model and a function that maps a random normal distribution to the parameterization space.
To this end, we rely on the Latent Space Model (LSM)~\cite{Wei23a, Wei23b} to provide an automatic shape parameterization represented by a learned latent space, denoted as $\bZ$.
Then we introduce the Latent Space Diffusion Model (LSDM, denoted as $g$), which progressively denoises a random normal vector into a valid latent code.

\subsection{Latent Space Geometry Parameterization}
For the latent representation, we employ the LSM to parameterize the design surface geometry. It relies on an auto-decoder design \cite{Tan95, Park19c} and is trained jointly with the latent space using a dataset of the collected shapes. This model maps a template shape $\hat{M}=\{\hat{V}, E\}$ with given $N$ vertices $\hat{V}=\{\bv_1, \bv_2,...,\bv_N\}$ and edges $E$, such as NACA-0012 for 2D airfoils or a mean shape in a more general case, conditioned on a $d$ dimensional latent vector $\bz$ that parameterizes the output geometry as
\begin{align}
    f_{\Theta} : \mathbb{R}^d \times \mathbb{R}^D \rightarrow \mathbb{R}^D \; , \label{eq:map_lsm} \\
    \mathbf{\delta v} = f_{\Theta}(\bz, \hat{\bv}) \; , \nonumber
\end{align} 
where $D\in {2,3}$ stands for the coordinate space dimensionality and 
\revise{$\delta \bv$ for the vertex displacement.}
LSM is a multi-layer perceptron (MLP) network and $\Theta$ represents the weights that control LSM $f$.
A deformed shape is obtained as $M=\{\hat{V} + f_{\Theta}(\bz, \hat{V}), E\}$. 

During training, we collect a dataset of $K$ surface geometries, denoted as $S_1,\ldots,S_K$, which only contains sampled surface points. The training objective is to optimize the weights $\Theta$ and the latent vectors that parameterize each training data $\textbf{z}_1,\ldots,\textbf{z}_K$. We write this as:

\revise{
    \begin{align}
      \Theta^*,\bZ^* 
        &=  \argmin_{\Theta,\bz_1,\ldots,\bz_K} \sum_{k=1}^K \cL_{LSM}(\hat{V} + f_{\Theta}(\bz_k, \hat{V}), S_k)\;,
    \label{eq:argmin_f_z_theta}
    \end{align}
}
where $\cL_{LSM}$ is a loss function that is minimized when $f_{\Theta}(\bz_k, \hat{V})$ yields a deformed airfoil geometrically identical to $S_k$. The optimal $\bz_k^*$ corresponds to a low-dimensional parameterization of $S_k$. We use the chamfer distance~\cite{Barrow77}, denoted as $\cL_{CD}(V, S)$, to measure the geometric difference.  Additionally, we apply a vector norm regularization on $\bz$ to ensure the smoothness of the acquired latent space. Therefore, the overall training objective becomes:
\begin{align}
  \cL_{LSM}(V, S) 
    &= \cL_{CD}(V, S) + w_{\bz}\left|\left|\bz\right|\right|^2 \nonumber \\
    &= \underbrace{\sum_{v\in V}\min_{s\in S}\left\|v - s\right\|_2^2\;+\;\sum_{s\in S}\min_{v\in V}\left\|s - v\right\|_2^2}_{\cL_{CD}(V,S)}  + w_{\bz}\left|\left|\bz\right|\right|^2 \; ,
\end{align}
where $w_{\bz}$ is the balancing weight.

At inference time, given a new target geometry $S$ and the frozen weights $\Theta^*$, the latent parameterization vector $\bz^*$ is obtained by finding
\revise{
    \begin{align}
      \bz^* &=  \argmin_{\bz} \cL_{LSM}(\hat{\bv} + f_{\Theta^*}(\bz, \hat{\bv}), S) \; .
    \label{eq:agmin_f_z}
    \end{align}
}
The output geometry becomes $M=\{\hat{V} + f_{\revise{\Theta^*}}(\bz^*, \hat{V}), E\}$. 

To synthesize new geometries, we simply need to create new latent codes that fall within the distribution of plausible latent vector and then decode it into a deformed shape. However, it is intractable to directly sample from the distribution of $\bZ$, which is unknown and non-analytical. Therefore, we require the LSDM to sample from a feasible random distribution and project the sampled random vector to a valid latent vector.

\subsection{Latent Space Diffusion Model for Unconditional Generation}

\begin{figure}[!t]
    \begin{center}
        \includegraphics[width=0.95\linewidth]{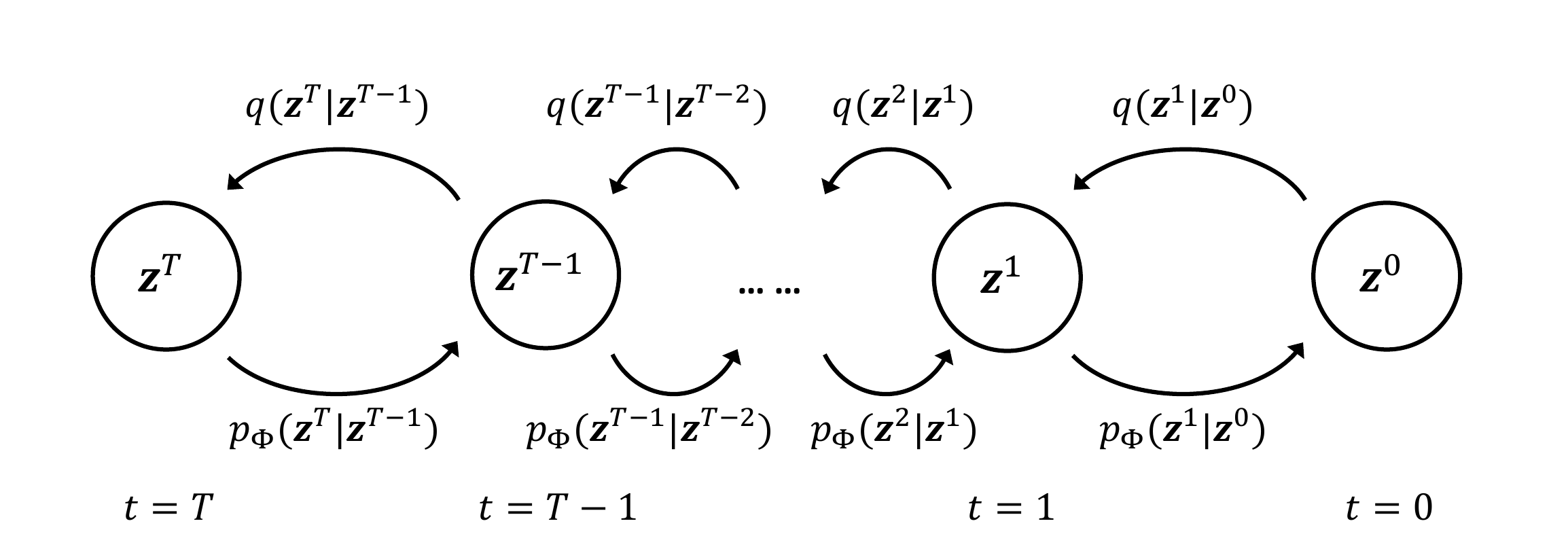}
    \end{center}
    \vspace{-4mm}
    \caption{
        \small \revise{LSDM's forward and reverse processes represented in latent-space Markov chain.}
    }
    \label{fig:abs_ddpm}
\end{figure}

The LSDM is a probabilistic generative model that maps a standard multivariate normal distribution to the latent space of the LSM via a Markov chain parameterization. It follows the denoising diffusion probabilistic model (DDPM) framework~\cite{Ho20a}, which defines a forward noising process $q$, and a learned reverse process $p_\Phi$, as shown in Fig.~\ref{fig:abs_ddpm}(a). For clarity, Fig.\ref{fig:abs_ddpm}(b) illustrates an analogous image-based diffusion process.

In the forward process, Gaussian noise is added with variance schedules $\{\beta_t\}^T_{t=1}$, $\alpha_t=1-\beta_t$ and $\bar{\alpha}_t=\Pi^t_{s=1}\alpha_s$. The noisy latent vector at step $t$ can be efficiently sampled in closed form as:
\begin{equation}
    q(\bz^t | \bz^0) = \cN(\bz^t; \sqrt{\Bar{\alpha}_t} \bz^0, (1-\Bar{\alpha}_t) \bI)\;.
\end{equation}
The reverse process starts from $p(\bz^T)=\cN(\bz^T; \textbf{0}, \bI)$ and uses Gaussian transitions $p_\phi(\bz^{t-1} | \bz^t) := \cN(\bz^{t-1}: \bmu_\phi(\bz^t, t), \beta_t \bI)$, where the mean is parameterized in the noise-prediction form:
\begin{equation}
    \bmu_\phi(\bz^t, t) := \frac{1}{\sqrt{\alpha_t}} (\bz^t - \frac{1-\alpha_t}{\sqrt{1-\Bar{\alpha}_t}}) g_\Phi(\bz^t, t)\;,
\end{equation}
where $g_\Phi(\bz^t, t)$ is the LSDM network based on the Multi-Layer Perceptron (MLP) structure.

The network is trained with the standard noise-matching loss:
\begin{equation}
    \cL_{LSDM} := \mathbb{E}_{t, \bz^0, \mathbf{\epsilon}\sim \cN(\textbf{0}, \bI)} 
    \left[ \left| \left|
    \mathbf{\epsilon} - g_\Phi \left(\sqrt{\Bar{\alpha}_t} \bz^0 + \sqrt{1-\Bar{\alpha}_t}\mathbf{\epsilon}, t \right)
    \right|\right| ^ 2\right]\;,
\end{equation}
with training steps summarized in Algorithm~\ref{alg:abs_train_diffusion}.
\begin{algorithm}
    \caption{The training steps of the diffusion network in LSDM.}
    \label{alg:abs_train_diffusion}
    \begin{algorithmic}
        \While {not converged}
            \State sample a latent code from the training dataset as $\bz^0$.
            \State $t \sim \text{Uniform}(\{1,2,...,T\})$ 
            \State $\mathbf{\epsilon} \sim \cN(\mathbf{0}, \bI)$
            \State Take gradient descent optimization step on $\nabla_\Phi \left|\left| \mathbf{\epsilon} - g_\Phi \left(\sqrt{\Bar{\alpha}_t} \bz^0 + \sqrt{1-\Bar{\alpha}_t}\mathbf{\epsilon}, t \right) \right|\right|^2$.
        \EndWhile
    \end{algorithmic}
\end{algorithm}

From the score-matching theory~\cite{song2021scorebased}, the trained $g_\Phi$ estimates the score function of the reverse process: 
\begin{equation}
    \nabla_{\bz^t} \log p_\Phi(\bz^t|\bz^{t+1})=-\frac{1}{\sqrt{1-\Bar{\alpha}_t}}  g_\Phi(\bz^t, t) \;.
\end{equation}
Unconditional sampling starts from $\bz^T\sim \cN(\textbf{0}, \bI)$ and iteratively applies the reverse transition (summarized in Algorithm~\ref{alg:abs_sample_diffusion}) with the stochastic gradient Langevin dynamics~\cite{welling11} until $\bz^0$ is obtained and decoded to a shape.

\begin{algorithm}
    \caption{The unconditional sampling steps of LSDM.}
    \label{alg:abs_sample_diffusion}
    \begin{algorithmic}
        \State $\bz^T \sim \cN(\mathbf{0}, \bI)$
        \For{$t=T,T-1,...,1$}
            \State $\mathbf{\epsilon} \sim \cN(\mathbf{0}, \bI)\;\text{if } t>1\;\text{, else } \mathbf{\epsilon}=0$ 
            \State $\bz^{t-1} = \frac{1}{\sqrt{\alpha_t}} \left(\bz^t - \frac{1-\alpha_t}{\sqrt{1-\Bar{\alpha}_t}} g_\Phi(\bz^t, t) \right) + \sqrt{\beta_t}\mathbf{\epsilon}$
        \EndFor 
        \State \textbf{return} {$\bz^0$}
    \end{algorithmic}
\end{algorithm}

\subsection{Conditional Latent Space Diffusion Model}
When developing generative models for aerospace engineering design, the sampling should be biased toward regions that satisfy design objectives and constraints. Formally, this amounts to solving
\begin{align}
    & \min_{V} O(V) \; , \nonumber\\
    \text{subject to} \quad & C^E_i(V) = 0 \;\; (i=1,\dots,m) \; , \\
    & C^I_j(V) \le 0 \;\; (j=1,\dots,r) \; , \nonumber
\end{align}
where $O$ denotes the primary objective, $C^E_i$ are equality constraints and $C^I_j$ are inequality constraints. $m$ and $r$ are the amounts of each type of constraint.
To impose these constraints during generation, we define an energy function that quantifies constraint satisfaction as a differentiable scalar loss. This energy function takes the form of a penalty:
\begin{equation}
    E(V)=\bigl\|O(V)\bigr\|^2 + \sum_{i=1}^m \rho_i \bigl\|C^E_i(V)\bigr\|^2 + \sum_{j=1}^r \gamma_j \bigl\|\max(0,C^I_j(V))\bigr\|^2 \quad ,
    \label{eq:energy_guidance}
\end{equation}
where $\rho_i > 0, \gamma_j > 0$ are penalty coefficients weighting the violation of equality and inequality constraints, respectively.

We incorporate this energy function into the sampling process as a guiding signal and formulate the conditional latent-space diffusion model (CLSDM). At each reverse diffusion step $t$, the energy is evaluated based on the current latent code $\bz^t$ and the decoded geometry $V^t=\hat{V}+f_\Theta(\bz^t)$. For simplicity, we denote: $E(\bz^t):=E(f_\Theta(\bz^t))=E(V^t)$.

To bias the sampling toward low-energy geometries, we define an unnormalized energy-augmented distribution over $\bz^t$, which we write as:
\begin{equation}
    \hat{p}_\Phi(\bz^t) \propto p_\Phi(\bz^t)e^{-\xi E(\bz^t)},
\end{equation}
where $\xi>0$ is a temperature parameter controlling the strength of guidance. At the sampling timestep $t$, The adjusted score function under this distribution becomes:
\begin{equation}
    \nabla_{\bz^t} \log \hat{p}_\Phi(\bz^t|\bz^{t+1})=-\frac{1}{\sqrt{1-\Bar{\alpha}_t}}  g_\Phi(\bz^t, t) - \xi\nabla_{\bz^t}E(\bz^t)\;.
\end{equation}

This modification augments the original score function of reversed process with a gradient that encourages geometries satisfying the design objectives. The full conditional sampling procedure is outlined in Algorithm~\ref{alg:abs_sample_conditional_diffusion}.
\begin{algorithm}
    \caption{The conditional sampling steps of CLSDM.}
    \label{alg:abs_sample_conditional_diffusion}
    \begin{algorithmic}
        \State Given: generation target in differentiable forms of $O,C^E,C^I$.
        \State $\bz^T \sim \cN(\mathbf{0}, \bI)$
        \For{$t=T,T-1,...,1$}
            \State $\mathbf{\epsilon} \sim \cN(\mathbf{0}, \bI)\;\text{if } t>1\;\text{, else } \mathbf{\epsilon} \text{ is all zero}$ 
            \State Compute $E(\bz^t)$,
            \State $\bz^{t-1} = 
                \frac{1}{\sqrt{\alpha_t}} 
                \left( 
                    \bz^t - 
                    \frac{1-\alpha_t}{\sqrt{1-\Bar{\alpha}_t}} g_\Phi(\bz^t, t) - 
                    \xi\nabla_{\bz^t} E(\bz^t) 
                \right) 
                + \sqrt{\beta_t}\mathbf{\epsilon}$
        \EndFor 
        \State \textbf{return} {$\bz^0$}
    \end{algorithmic}
\end{algorithm}

When the energy function encodes complex design objectives or tight constraints, especially in 3D shape generation, the two goals of sampling from the learned distribution and minimizing the energy function may not align perfectly in convergence speed within the fixed number of diffusion steps $T$. As a result, the final samples may not fully satisfy the design criteria.

To address this issue, we propose an enhanced conditional sampling strategy for CLSDM that extends Algorithm~\ref{alg:abs_sample_conditional_diffusion}. After the initial generation, if the decoded geometry fails to meet the desired objectives, we partially reintroduce noise to the latent vector by simulating a forward diffusion process up to an intermediate noise level $T^*$, where $1<T*<T$. The reverse diffusion is then restarted from this noisy latent vector. This re-noising and re-generation loop is repeated for a fixed number of iterations, allowing the model to explore alternative paths while remaining near the solution manifold. The complete procedure of enhanced conditional sampling is summarized in Algorithm~\ref{alg:abs_sample_enhanced_conditional_diffusion}.

\begin{algorithm}
    \caption{The enhanced conditional sampling steps of CLSDM.}
    \label{alg:abs_sample_enhanced_conditional_diffusion}
    \begin{algorithmic}
        \State Given: generation target in differentiable forms of $O,C^E,C^I$;
        \State \qquad\quad number of enhanced loops $N_{T^*}$;
        \State \qquad\quad noising level $T^*$.
        \State $\bz^T \sim \cN(\mathbf{0}, \bI)$
        \For{$t=T,T-1,...,1$}
            \State $\mathbf{\epsilon} \sim \cN(\mathbf{0}, \bI)\;\text{if } t>1\;\text{, else } \mathbf{\epsilon}\text{ is all zero}$ 
            \State Compute $E(\bz^t)$,
            \State $\bz^{t-1} = 
                \frac{1}{\sqrt{\alpha_t}} 
                \left( 
                    \bz^t - 
                    \frac{1-\alpha_t}{\sqrt{1-\Bar{\alpha}_t}} g_\Phi(\bz^t, t) - 
                    \xi\nabla_{\bz^t} E(\bz^t) 
                \right) 
                + \sqrt{\beta_t}\mathbf{\epsilon}$
        \EndFor 
        \For{$n=1,2,...,N_{T^*}$}
            \State $\mathbf{\tau} \sim \cN(\mathbf{0}, \bI)$
            \State $\bz^{T^*} = \sqrt{\Bar{\alpha}_{T^*}} \bz^0 + \sqrt{1-\Bar{\alpha}_{T^*}}\mathbf{\tau}$
            \For{$t=T^*,T^*-1,...,1$}
                \State $\mathbf{\epsilon} \sim \cN(\mathbf{0}, \bI)\;\text{if } t>1\;\text{, else } \mathbf{\epsilon}=0$ 
                \State Compute $E(\bz^t)$,
                \State $\bz^{t-1} = 
                    \frac{1}{\sqrt{\alpha_t}} 
                    \left( 
                        \bz^t - 
                        \frac{1-\alpha_t}{\sqrt{1-\Bar{\alpha}_t}} g_\Phi(\bz^t, t) - 
                        \xi\nabla_{\bz^t} E(\bz^t) 
                    \right) 
                    + \sqrt{\beta_t}\mathbf{\epsilon}$
            \EndFor
        \EndFor
        \State \textbf{return} {$\bz^0$}
    \end{algorithmic}
\end{algorithm}

\subsection{Implementation Details}
Both LSM and LSDM are implemented using the PyTorch~\footnote{pytorch.org} toolbox and are capable of GPU acceleration. The dimension of LSM's latent space is $d=256$. \revise{This dimension was selected empirically to ensure sufficient capacity for high-fidelity reconstruction while maintaining computational efficiency across different cases.} We use the Adam optimizer~\cite{Kingma15a} with the learning rate of $5\times10^{-4}$ for LSM's weights $\Theta$ and $10^{-3}$ for the latent space $\bZ$. \revise{The weight regularization parameter is $w_z=10^{-4}$. The LSM decoder is a 4-layer MLP with hidden widths of $[256, 256, 512, 512]$.}

LSDM is implemented as \revise{a 5-layer MLP with hidden widths of $[256, 512, 512, 512, 256]$} and leaky ReLU activation functions. The forward and reverse processes use $T=1000$ time steps. The diffusion scheduler $\beta$ linearly increases from $10^{-4}$ to $0.02$ as $t$ ranges from $0$ to $T$, and remains constant in both forward and reverse processes. The LSDM is trained with the AdamW optimizer~\cite{Loshchilov19} and a learning rate of $10^{-5}$. \revise{For conditional sampling, the guidance temperature $\xi$ follows a time-dependent schedule scaled by the noise variance $\beta_t$ (e.g., $\xi=10^6 \beta_t$ for area constraints). When applying the enhanced conditional sampling, we utilize $N_T=3$ loops with a re-noising level of $T^*=300$.} Under these implementation settings, \textit{DiffGeo} is computationally efficient. For example, when developing the 2D airfoil generator, the training of LSM and LSDM with a dataset of 50 airfoils take $671.9$ seconds and $716.4$ seconds on an NVIDIA V100 graphics card, respectively.
The averaged time cost of unconditional airfoil generation is $690.51$ milliseconds per 50 samples.
\section{Experiments and Results}
\label{sec:exp}

To investigate \textit{DiffGeo}’s capabilities across a range of aerodynamic design tasks, we perform three case studies in both 2D and 3D contexts. These experiments are designed to evaluate the method’s effectiveness in representative scenarios that reflect practical design challenges:
\begin{itemize}
    \item \textbf{Sampling Model Benchmark}: We compare \textit{DiffGeo} with GAN- and VAE-based generative models on a 2D airfoil generation task. The comparison focuses on sampling quality, novelty, diversity and constraint adherence, especially under data-limited settings.

    \item \textbf{Generating Task-Informed Training Data for Surrogate Models}: We integrate \textit{DiffGeo} into a surrogate-based optimization (SBO) pipeline for airfoil design, using conditional sampling to embed design objectives as priors when generating training data for the surrogate models.

    \item \textbf{3D Design Prototyping}: We assess whether \textit{DiffGeo} can produce feasible 3D blade geometries under realistic data scarcity while conforming to high-dimensional design constraints. We also investigate its potential to enhance conventional design tools by automating and accelerating 3D blade prototyping.
\end{itemize}

These case studies address three key research questions:
\begin{itemize}
    \item[1.] \textbf{Data Efficiency}: Can \textit{DiffGeo} support high-quality, novel and controllable design space exploration with only a limited amount of training data?

    \item[2.] \textbf{Deployment Flexibility}: Does decoupling geometry generation from design objectives allow the generative model to be reused across different design tasks without retraining?

    \item[3.] \textbf{Constraint Handling}: To what extent can \textit{DiffGeo} integrate complex, fine-grained and high-dimensional constraints into the generative process?
\end{itemize}
\subsection{Benchmarking Generative Models on 2D Airfoils}
\label{sec:benchmarking}

Generative adversarial networks (GANs) and variational autoencoders (VAEs) represent two state-of-the-art frameworks for data-driven aerodynamic design space exploration. To benchmark \textit{DiffGeo}’s diffusion-based approach, we compare it against these baselines on a common task of 2D airfoil generation. Importantly, all models operate within the same learned shape parameterization: the automatic latent space model (LSM) introduced by ~\citet{Wei23a,Wei23b}. The advantages of this automatic LSM have been thoroughly discussed in prior work. Using a shared latent representation ensures a fair comparison by isolating the effect of the sampling mechanism from differences in geometry encoding. We examine three key aspects of generative performance:

\begin{itemize}
    \item \textbf{LSDM vs. Adversarial Training}: Both \textit{DiffGeo} and the GAN baseline generate airfoils by sampling in the latent space learned by the LSM. The GAN uses adversarial training to produce latent codes, while \textit{DiffGeo} uses a latent space diffusion model (LSDM) for more stable and data-efficient sampling. We compare how these approaches perform, particularly when training data is extremely limited.

    \item \textbf{Generative Sampling vs. VAE Prior}: Unlike LSDM and GANs which rely on explicit external samplers, a VAE-style model integrates sampling by learning a Gaussian prior in the latent space. For comparison, we adapt the LSM into a variational auto-decoder (VAD) to enforce a learned latent Gaussian distribution (similar to a VAE) and evaluate whether this prior-based sampling strategy affects output diversity and quality differently than LSDM’s learned sampler.

    \item \textbf{Energy-Based Guidance vs. Conditional GAN}: \textit{DiffGeo}’s conditional extension CLSDM uses energy-based guidance during sampling to incorporate constraints without retraining the generator. In contrast, a conditional GAN (CGAN) baseline requires paired data and retraining to implement new conditions. We assess each model’s ability to satisfy design constraints under limited data--comparing CLSDM’s flexible, plug-and-play guidance to CGAN’s fixed, learned conditioning.
\end{itemize}

\subsubsection{Generative Adversarial Network Baseline}
For the GAN baseline, we implement a generator–discriminator pair that shares the LSM latent space with \textit{DiffGeo}. 

The generator $g_{\Gamma}^{(G)}: \mathbb{R}^d \to \bZ$ is a multilayer perceptron (MLP) that maps a random Gaussian noise $\epsilon \sim \cN(\mathbf{0},\bI)$ to a latent code $\bz = g^{(G)}_\Gamma(\mathbf{\epsilon})$. It uses the same architecture as \textit{DiffGeo}’s diffusion network, but is trained via an adversarial objective. The discriminator $g_{\Pi}^{(D)}: \bZ \rightarrow [0,1]$ is an MLP that attempts to distinguish `real' latent codes $\bz_{\text{real}}$ obtained by auto-decoding from training data, from `fake' codes generated by $g^{(G)}_\Gamma$. It consists of two hidden layers, which is an empirical setting for stable adversarial training. Both networks are trained with the standard \textit{minimax} objective~\cite{Goodfellow14} using the AdamW optimizer and a learning rate of $10^{-5}$:
\begin{equation}
    \min_\Gamma \max_\Pi = 
    \mathbb{E}_{\hat{\bz}_\text{real} \sim p_{data}} \left[ \log g^{(D)}_\Pi(\hat{\bz}_\text{real}) \right] + 
    \mathbb{E}_{\mathbf{\epsilon} \sim \cN(\mathbf{0}, \bI)} \left[ \log \left( 1 - g^{(D)}_\Pi( g^{(G)}_\Gamma(\mathbf{\epsilon}) ) \right)\right]\;,
\end{equation}

After training, the GAN generator can produce new latent samples $\bz=g^{(G)}(\mathbf{\epsilon})$, which are then decoded by the LSM’s decoder to obtain a new airfoil $M=\{\hat{V} + f_{\Theta}(\bz, \hat{V}), E\}$, where $\bar{V}$ are template mesh vertices and $f_{\Theta}$ is the LSM's deformation function.

We also implement a conditional GAN (CGAN) variant for comparisons. The CGAN generator and discriminator are augmented to accept a conditioning input $C$ (e.g., a target parameter) as $g_{\Gamma}^{(G)}(\mathbf{\epsilon},C)$ and $g_{\Pi}^{(D)}(\bz,C)$. This allows GAN sampling under a given condition by learning through training on paired data, though it lacks the flexibility of \textit{DiffGeo}’s energy-based conditioning.

\subsubsection{Variational Auto-Decoder Baseline}
As a second generative baseline, we convert the LSM into a variational latent model similar to a VAE, but without an encoder network. Specifically, we treat each training shape $S_1,\dots,S_K$ as having its own latent distribution rather than a fixed code, thereby creating a variational auto-decoder (VAD). During training, we optimize the decoder’s weights $\Theta$ and learn a mean $\mathbf{\mu}_i\in \mathbb{R}^d$ and standard deviation $\mathbf{\sigma}_i\in \mathbb{R}^d_{>0}$ for the latent vector of each training shape $S_i$. This effectively enforces a Gaussian prior in the latent space: the learned $\{\mathbf{\mu}_i, \mathbf{\sigma}_i\}$ approximate the shape’s posterior, and a new latent code $\bz_i$ can be sampled by perturbing these distributions using the reparameterization trick:

\[\bz_k=\mathbf{\mu}_i+\mathbf{\sigma}_i\mathbf{\epsilon}, \quad \mathbf{\epsilon\sim \cN(\mathbf{0}, \bI)}\;.\]

The sampled latent code is passed to VAD's decoder $f^{V}_\theta$ to generate deformed airfoil. The model is trained to minimize the loss:
\begin{equation*}
    \min_{\Theta,\{\mathbf{\mu}_i, \mathbf{\sigma}_i\}} \;
  \sum_{i=1}^K \left[
    \cL_{CD}(\hat{V}+f^V_{\Theta}(\bz_i), S_i) 
    + \cD_{KL}\bigl(\cN(\mathbf{\mu}_i, \operatorname{diag}(\mathbf{\sigma}_i^2)) \,\|\, \cN(\mathbf{0}, \bI)\bigr)
  \right].
\end{equation*}
The chamfer distance $\cL_{CD}$ minimizes the reconstruction error, while the Kullback–Leibler divergence $\cD_{KL}$ regularizes the latent space toward a standard normal prior. The reparameterization allows gradients to flow through stochastic sampling and enables end-to-end training.

We use the same decoder architecture as the original LSM, training $\Theta$ with a learning rate of $5\times10^{-4}$, while the latent distribution parameters ${\mathbf{\mu}_i,\mathbf{\sigma}_i}$ are updated with a higher learning rate $10^{-3}$. The VAD baseline integrates the sampling mechanism into the model via the learned Gaussian priors, providing a point of comparison for \textit{DiffGeo}’s external sampler LSDM versus an internal latent prior approach.

\subsubsection{Data Preparation}
We collect airfoils from the UIUC database~\cite{Selig96a}, which offers a wide variety of foil profiles. \revise{Prior to training, the database was manually filtered to remove noisy or incomplete shapes, and all profiles were normalized to a unit chord length.} \revise{Airfoils for LSM training were interpolated to $N=600$ points along the contour. The template shape $\hat{M}$ used for the 2D case is the NACA-0012 discretized into 200 points.} Five subsets are randomly sampled, each consisting of $1000$, $500$, $250$, $100$, and $50$ airfoils, respectively.

\begin{figure}[!t]
    \begin{center}
        \includegraphics[width=1\linewidth]{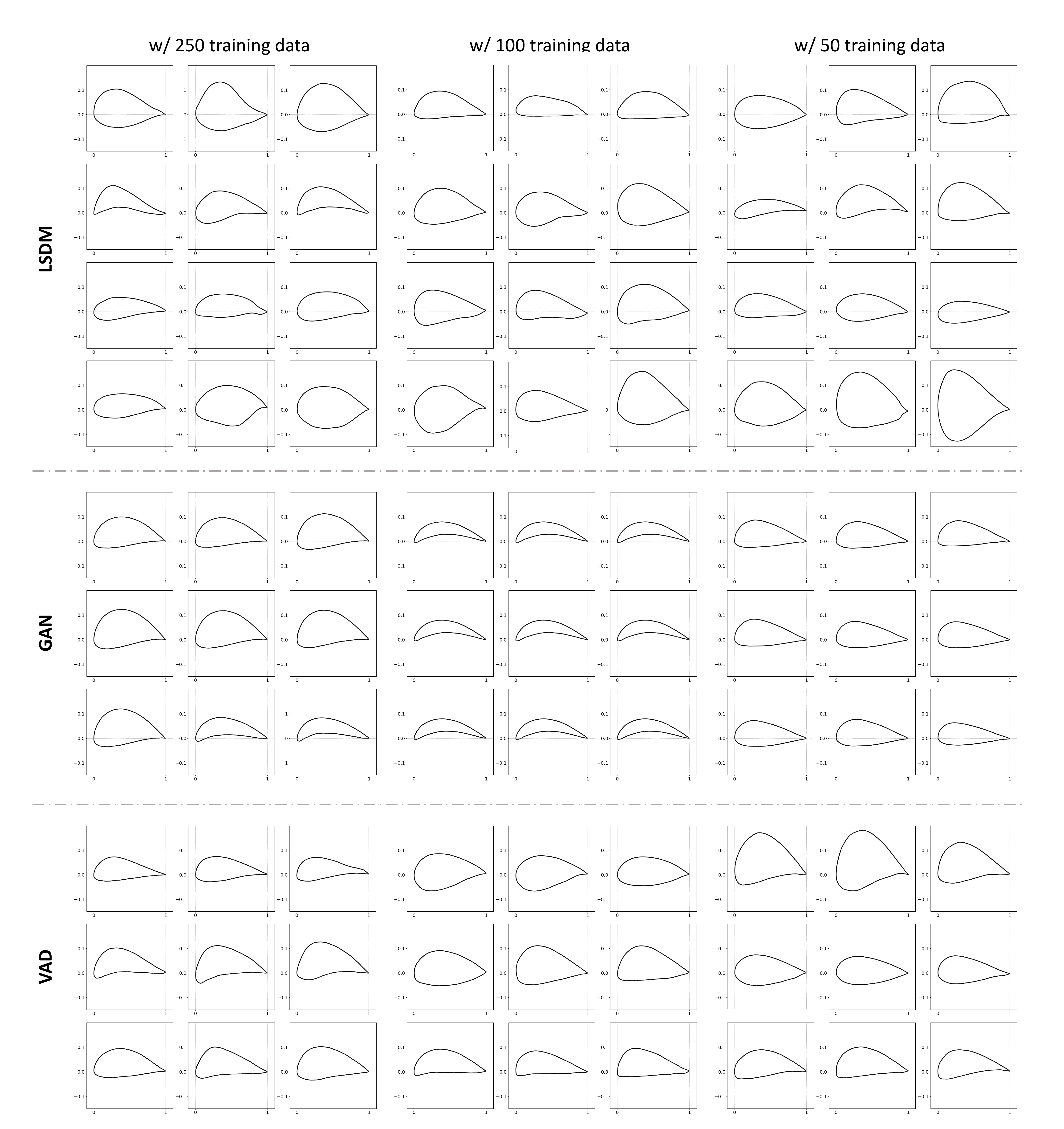}
    \end{center}
    \caption{
        \small Unconditional airfoil samples generated by GAN, VAD and LSDM.
    }
    \label{fig:main_unconditional_airfoils}
\end{figure}

\begin{figure}[!t]
    \begin{center}
        \includegraphics[width=0.95\linewidth]{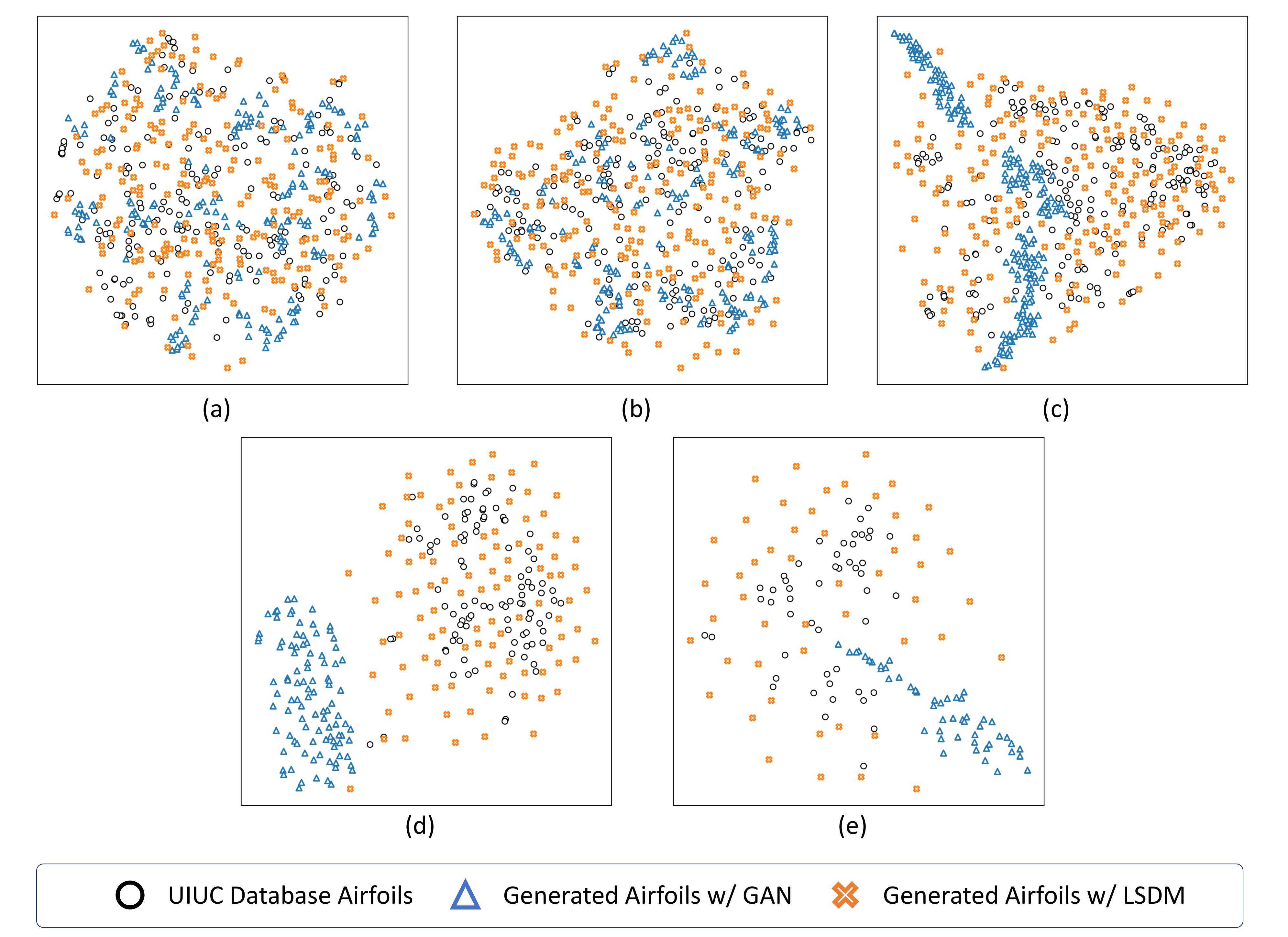}
    \end{center}
    \caption{
        \small \textit{t-SNE} visualization of airfoil latent codes across five training set sizes (a) $1000$, (b) $500$, (c) $250$, (d) $100$, (e) $50$ from UIUC database, LSDM and GAN.
    }
    \label{fig:main_unconditional_tsne}
\end{figure}

\begin{figure}[!h]
    \begin{center}
        \includegraphics[width=1\linewidth]{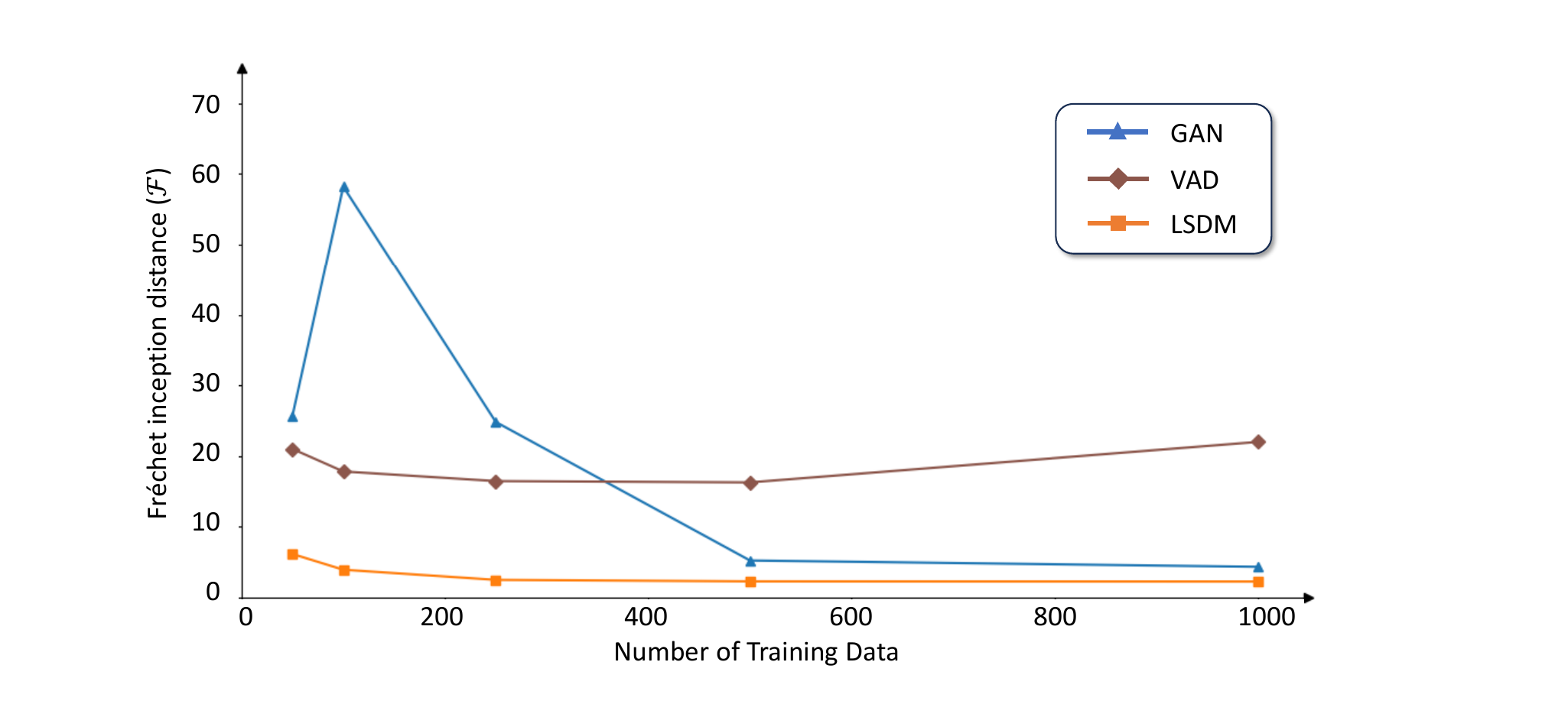}
    \end{center}
    \vspace{-4mm}
    \caption{
        \small FID scores of airfoils generated by GAN, VAD and LSDM across varying training set sizes. 
    }
    \label{fig:main_benchmark_fid}
\end{figure}

\subsubsection{Qualitative Comparison}

Fig.~\ref{fig:main_unconditional_airfoils} provides a visual comparison of airfoils generated by each method (GAN, VAD and LSDM) with progressively smaller training sets. All models produce smooth airfoils with their chords aligned, which reflects the strong geometric prior enforced by the LSM representation. However, there are clear differences in diversity. LSDM generates airfoils with a wide range of shapes with significant variations in thickness distribution, camber, leading-edge radius and position of maximum thickness. This variability still exists even when trained on as few as 50 airfoils. By contrast, GAN suffers from mode collapse given limited data. With fewer than 250 training samples, many GAN outputs collapse to nearly identical shapes and the overall shape variety nearly disappears. VAD does not exhibit catastrophic mode collapse, but its outputs tend to cluster into a few modes. The generated airfoils are less diverse than LSDM’s results and often fall into several families of similar shapes. These qualitative trends indicate that \textit{DiffGeo}’s diffusion-based sampler better preserves shape diversity even under limited data, whereas the GAN struggles and the VAE-style approach produces only moderate diversity.

Fig.\ref{fig:main_unconditional_tsne} visualizes the latent space coverage of LSDM and GAN using the \textit{t}-Distributed Stochastic Neighbor Embedding algorithm (\textit{t-SNE})~\cite{Maaten08}. In each subplot, grey points represent latent codes of dataset airfoils from the UIUC database, while orange and blue points represent latent codes generated by LSDM and GAN respectively for different numbers of training data from 1000 down to 50. We can observe that LSDM’s samples closely overlap the distribution of real airfoils in the latent space. Even with very few training shapes, LSDM can still effectively project the multivariate normal prior into LSM's learned manifold, so that the generated latent codes form a similar region as the real data. In contrast, GAN’s samples show significant distributional gaps and clustering. The GAN fails to cover large portions of the latent space that correspond to valid airfoils, and many generated points get clustered, especially given fewer than 250 training samples. These clusters reflect severe mode collapse issues, where GAN produces limited variations. Qualitatively, the \textit{t-SNE} plots demonstrate that LSDM captures the training shape distribution with higher fidelity than the GAN in data-scarce settings. The VAD outputs, though not shown in \textit{t-SNE} due to its probabilistic nature, were observed to generate clear clusters as well, indicating a restricted coverage of the latent space compared to \textit{DiffGeo}.

\subsubsection{Quantitative Evaluation on Sampling Quality}

To quantify how indistinguishable the generated shapes are from the real ones, we compute a Fréchet Inception Distance (FID, denoted as $\cF$) metric with adaptation to aerodynamic shapes. Instead of using an image classifier as in original FID, we use a CFD-based surrogate model trained to predict lift and drag coefficients of airfoils in UIUC database~\cite{Baque18} to extract feature vectors for each shape. \revise{Specifically, we utilize the 128-dimensional feature vector extracted from the last hidden layer of the surrogate model (a Geodesic Convolutional Neural Network ). This deep layer encodes high-level geometric features relevant to aerodynamic performance, ensuring the metric differentiates shapes based on meaningful structural properties rather than trivial coordinate noise.} We then calculate FID between the set of generated airfoils and the training dataset in this learned feature space. Given a set of airfoils, feature vectors are inferred from the last hidden layer of the surrogate model. By computing the mean feature vector $\mathcal{M}$ and covariance matrix $\Sigma$, the FID score between two sets of airfoils is defined as:
\begin{equation}
    \cF(\mathcal{M}_1, \mathcal{M}_2, \Sigma_1, \Sigma_2) = \left[ \left|\left| \mathcal{M}_1 - \mathcal{M}_2 \right|\right|^2 + \text{tr}\left( \Sigma_1 + \Sigma_2 - 2\sqrt{\Sigma_1\Sigma_2} \right) \right]^2_F\;.
\end{equation}
To quantify distributional gaps from generated airfoils and training set, $\mathcal{M}_1$ and $\Sigma_1$ are fixed from the UIUC database, while $\mathcal{M}_2$ and $\Sigma_2$ are computed from shapes generated by GAN, VAD and LSDM, respectively.

Fig.~\ref{fig:main_benchmark_fid} shows the FID scores as a function of training data size. LSDM achieves the lowest and best FID in all cases, indicating its generated airfoils are statistically closest to the distribution of training shapes. With very small training sets (e.g., 50 airfoils), GAN’s FID is much higher, reflecting that GAN outputs deviate significantly from the true data distribution. VAD shows intermediate performance as its FID is higher than LSDM’s but more stable than GAN’s with respect to different dataset. Notably, with larger training sets (e.g., more than 500 samples), GAN’s FID improves and approaches LSDM’s one but LSDM still keeps a slight advantage. In the extreme low-data scenario, LSDM dramatically outperforms GAN while LSDM’s generation fidelity remains largely intact. This highlights \textit{DiffGeo}’s robustness to data scarcity in terms of sample quality.

\begin{figure}[!t]
    \begin{center}
        \includegraphics[width=1\linewidth]{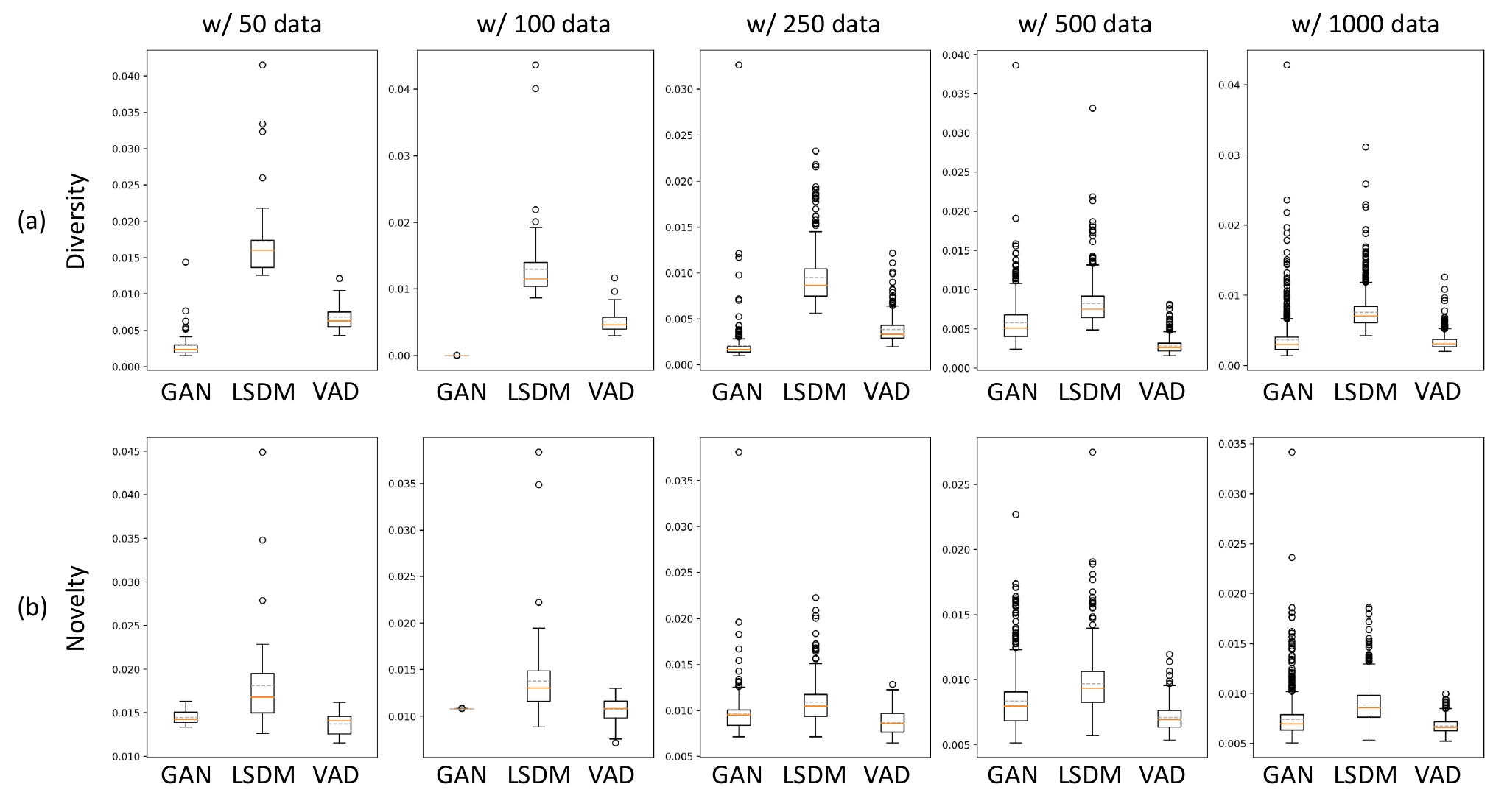}
    \end{center}
    \vspace{-4mm}
    \caption{
        \small Diversity ((a) $D_{intra}^{10}$) and novelty  ((b) $D_{inter}^{10}$) metrics for GAN, VAD and LSDM across dataset sizes. 
    }
    \label{fig:main_benchmark_unconditional_intra_inter_dist}
\end{figure}

\begin{figure}[!t]
    \begin{center}
        \includegraphics[width=1\linewidth]{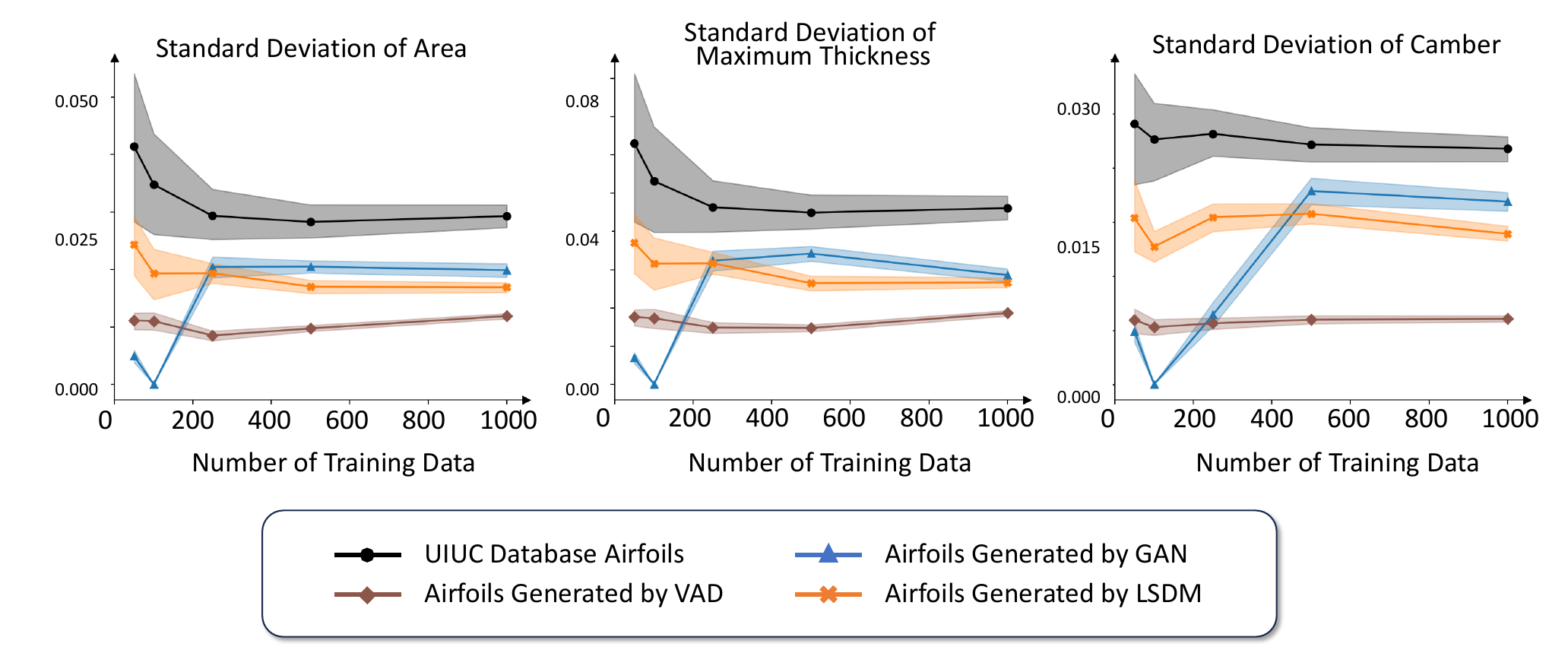}
    \end{center}
    \vspace{-2mm}
    \caption{
        \small The standard deviations of area, maximum thickness and camber for GAN, VAD and LSDM. 
    }
    \label{fig:main_benchmark_unconditional_std}
\end{figure}

\subsubsection{Quantitative Evaluation on Sampling Exploration}

A generative model that successfully explore the design space will generate diverse sets of novel designs. We assess this via two metrics: diversity among generated samples and novelty relative to the training set.

Diversity is a set metric, which measures model's generalization and the information entropy of generated samples. We define a $k$-nearest-intra-sample distance $D_{intra}^k$ to measure diversity. For each newly generated airfoil, we compute the average chamfer distance (CD) to its $k=10$ nearest neighbors within the generated set, then take the mean over all generated samples. A higher $D_{intra}^{10}$ means the set of outputs covers a broader range of shapes and have higher variability.

Novelty is a point metric that measures how far generated shapes deviate from real designs in the training set. Similarly, we define a $k$-nearest-inter-sample distance $D_{inter}^k$ as the average distance from each generated airfoil to its $k=10$ nearest neighbors in the training dataset. Higher values indicate the generator is producing shapes that are different from those in the training set while still remaining valid designs.

Fig.~\ref{fig:main_benchmark_unconditional_intra_inter_dist} demonstrates the computed $D_{intra}^{10}$ and $D_{inter}^{10}$ for GAN, VAD and LSDM under various training set sizes. LSDM has the highest scores in both diversity and novelty across all scenarios. With sufficient data (e.g., 1000 samples), all methods achieve certain diversity, but LSDM is still above GAN and VAD. As training data is reduced, GAN’s diversity and novelty collapse. For example, with 100 or 50 samples, GAN-generated airfoils show almost no diversity where $D_{intra}^{10}$ is close to zero and very low novelty, meaning that the few representative shapes GAN can generate are clustered near some training examples. VAD maintains a more consistent diversity and novelty level regardless of data size without any catastrophic mode collapse, but its absolute values are lower than LSDM in every case. In comparison, LSDM remains high diversity and novelty even with only 50 training airfoils and is not affected by data scarcity.

\begin{figure}[!th]
    \begin{center}
        \includegraphics[width=1\linewidth]{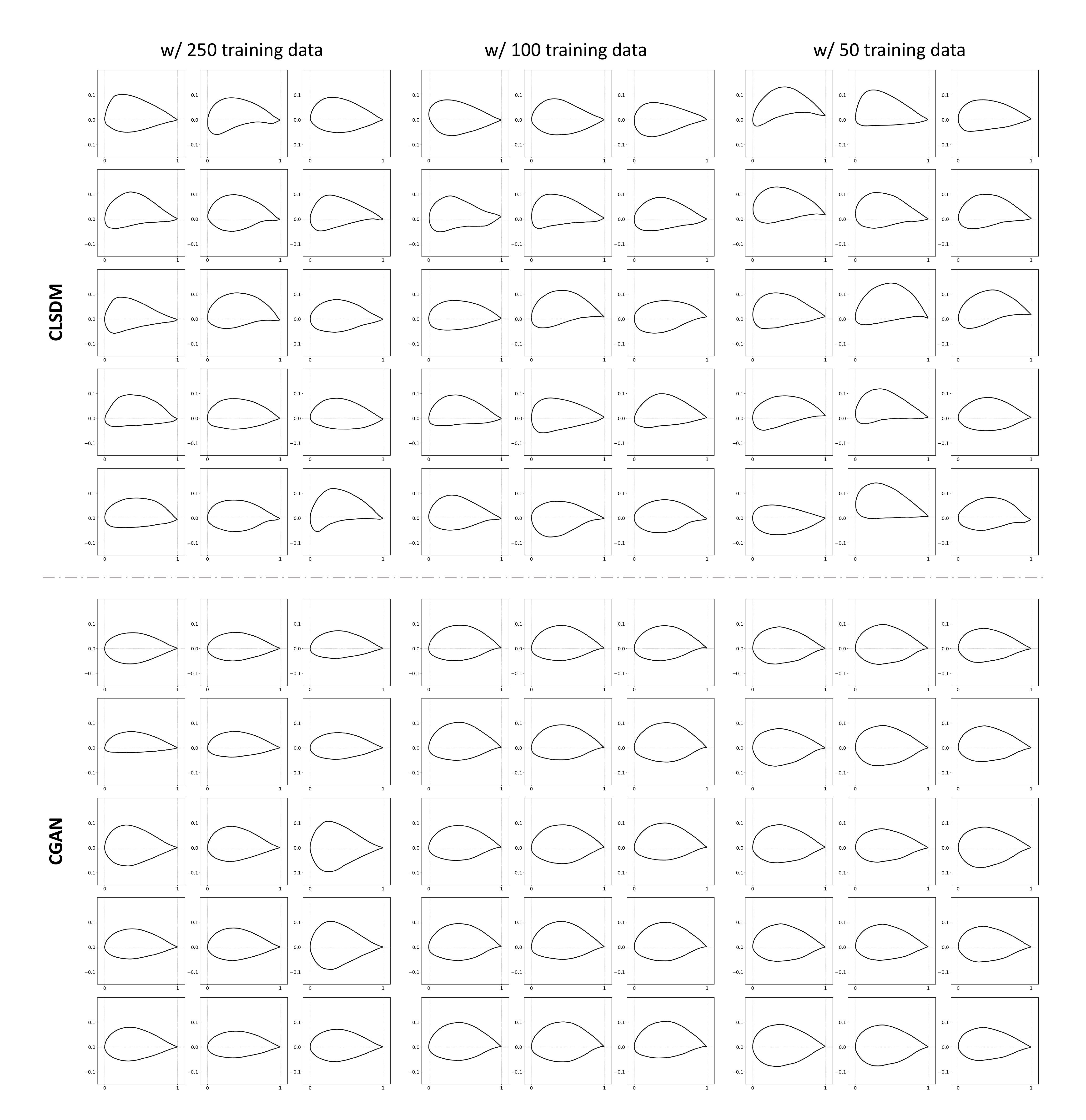}
    \end{center}
    \caption{
     \small Airfoils conditioned on $\mathcal{A}=0.09$ sampled by CLSDM and CGAN.
    }
    \label{fig:main_conditional_airfoils}
\end{figure}

We also directly measured the distribution of specific geometric properties among generated samples as shown in Fig.~\ref{fig:main_benchmark_unconditional_std}, including the standard deviations of airfoil area, maximum thickness and camber for each model. Consistent with the $D_{intra}^{10}$ metric, LSDM and VAD exhibit relatively flat variance curves as the number of training data varies, indicating they continue to produce a certain range of shapes. In contrast, GAN’s output variability drops with fewer than 200 training shapes. For very low data settings, GAN’s generated airfoils all have nearly the same area/thickness/camber, leading to negligible standard deviations that reflect severe mode collapse. VAD’s variance is more stable but is uniformly lower than that of LSDM, meaning that LSDM explores a wider range of geometries.

\subsubsection{Discussion}

In summary, the benchmark results demonstrate that \textit{DiffGeo}’s diffusion-based sampler LSDM consistently outperforms GAN- and VAE-based approaches in both sampling quality and sampling exploration, especially under low-data settings. Compared to adversarial training, LSDM offers stable generation without mode collapse, eliminating the need for large training sets or extensive hyperparameter tuning to stabilize GANs. Compared to the variational and prior-based sampling strategy, LSDM leverages the full representation capacity of the auto-decoder, which VAEs often compromise due to its regularization. As a result, LSDM produces more diverse and higher-fidelity samples. Although VAE and VAD can avoid mode collapse, their performance is degraded by a less expressive latent representation. In summary, for the unconditional generation task, \textit{DiffGeo} achieves nearly an order-of-magnitude improvement in data efficiency over GANs and shows clear advantages over VAD in diversity and novelty. These findings answer the first two comparison points: diffusion-based sampling performs better than adversarial training in stability, and than a fixed latent prior in flexibility and output quality.

\subsubsection{Conditional Generation Benchmark}
In this section,  we evaluate conditional generative performance by comparing the conditional LSDM (CLSDM) against conditional GAN (CGAN). This investigation addresses how well each approach can integrate design constraints into the generation process, especially with limited data. For a concrete comparison, we choose a simple conditioning objective, namely generating airfoils that satisfy a constant area value (e.g., $0.09$ for this case). The CGAN is retrained to model a joint probability of data and condition for each dataset size with area as the conditioning input, while CLSDM uses energy-based guidance to enforce the area during sampling without any additional training. 

To investigate conditional generation quality, we sample a large number of airfoils from each model under the condition of area being $0.09$. For CLSDM, we implement Equation~\ref{eq:energy_guidance} with the equality constraint $C^E$ that uses the closed-form Shoelace Formula to calculate the area $\mathcal{A}$. Given vertices on airfoil's contour in clockwise or anti-clockwise order $V=\{\bv_1, \bv_2,...\bv_N\}$, $\mathcal{A}$ is defined as:
\begin{equation}
    \mathcal{A}(V) = \frac{1}{2} \sum_{i=1}^N \left(v_{i,x}v_{i+1,y} - v_{i+1,x}v_{i,y} \right)\;,\;\text{where}\;\bv_i = (v_{i,x},v_{i,y})\;.
\end{equation}
Then we have $C^E = \bigl\| \mathcal{A}(V)-0.09\bigr\| ^2$. For CGAN, its input condition remains a constant scalar as $C=0.09$.

\begin{figure}[!t]
    \vspace{-6mm}
    \begin{center}
        \includegraphics[width=0.9\linewidth]{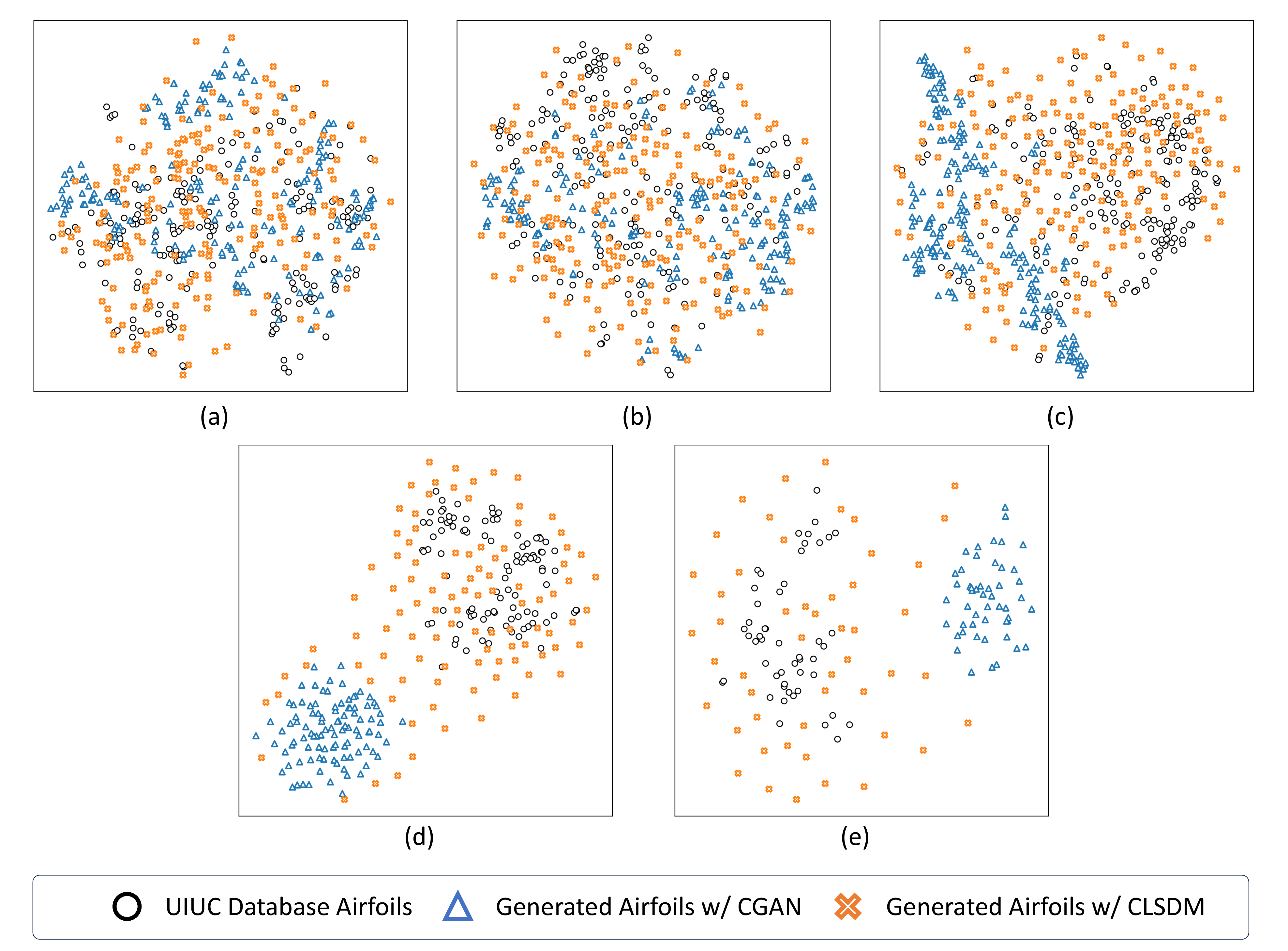}
    \end{center}
    \vspace{-4mm}
    \caption{
        \small \textit{t-SNE} visualization of airfoil latent codes across training set sizes (a) $1000$, (b) $500$, (c) $250$, (d) $100$ and (e) $50$ for UIUC database, CLSDM and CGAN when $\mathcal{A}=0.09$.
    }
    \label{fig:main_conditional_tsne}
\end{figure}

Fig.~\ref{fig:main_conditional_airfoils} shows generated airfoils sampled by CLSDM and CGAN. Similar as the conclusion draw from the unconditional generation benchmark, all samples exhibit valid airfoil properties, and CLSDM consistently demonstrates better diversity than CGAN, especially with fewer than 250 training data. Although airfoils sampled by CGAN trained with 250 samples show more variety, the area condition is not always respected, as can be observed in Fig.~\ref{fig:main_conditional_airfoils}(d). Fig.~\ref{fig:main_conditional_tsne} provides the \textit{t-SNE} visualization in the latent space. CLSDM aligns better with the training data distribution while CGAN's distributions are clearly biased and clustered, showing CGAN's poorer sampling quality and diversity compared to CLSDM.

\begin{figure}[!t]
    \begin{center}
        \includegraphics[width=1\linewidth]{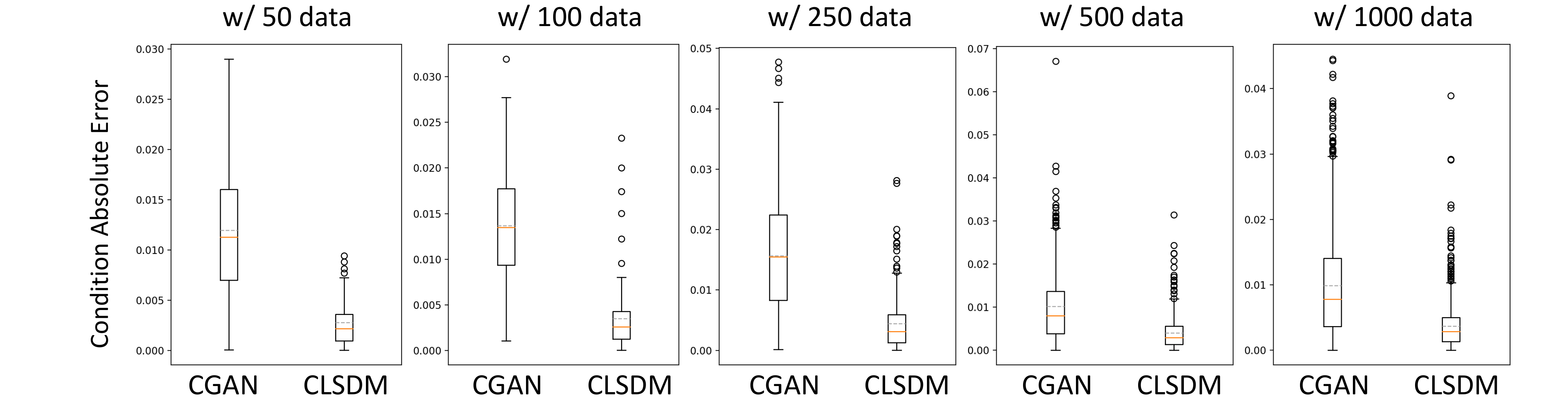}
    \end{center}
    \vspace{-2mm}
    \caption{
        \small Mean absolute area error of CLSDM and CGAN conditioned airfoils.
    }
    \label{fig:main_benchmark_condition_mae_boxplot}
\end{figure}

The condition adherence performance is measured by the difference between each generated airfoil’s actual area and the target value. Fig.~\ref{fig:main_benchmark_condition_mae_boxplot} plots the mean absolute error in area for CGAN and CLSDM at various training set sizes. CLSDM shows significantly lower error, indicating that \textit{DiffGeo} is able to satisfy the target area closely for its samples even with very few training data. Meanwhile, CGAN struggles to enforce the area constraint since its error is an order of magnitude higher and does not improve by a large margin with more data. This indicates that energy-based diffusion guidance is far more effective at satisfying constraints than the GAN’s learned conditioning, especially since the GAN must implicitly learn the mapping from condition to shape during training which is difficult with limited data and a single scalar condition.

\begin{figure}[!ht]
    \begin{center}
        \includegraphics[width=0.95\linewidth]{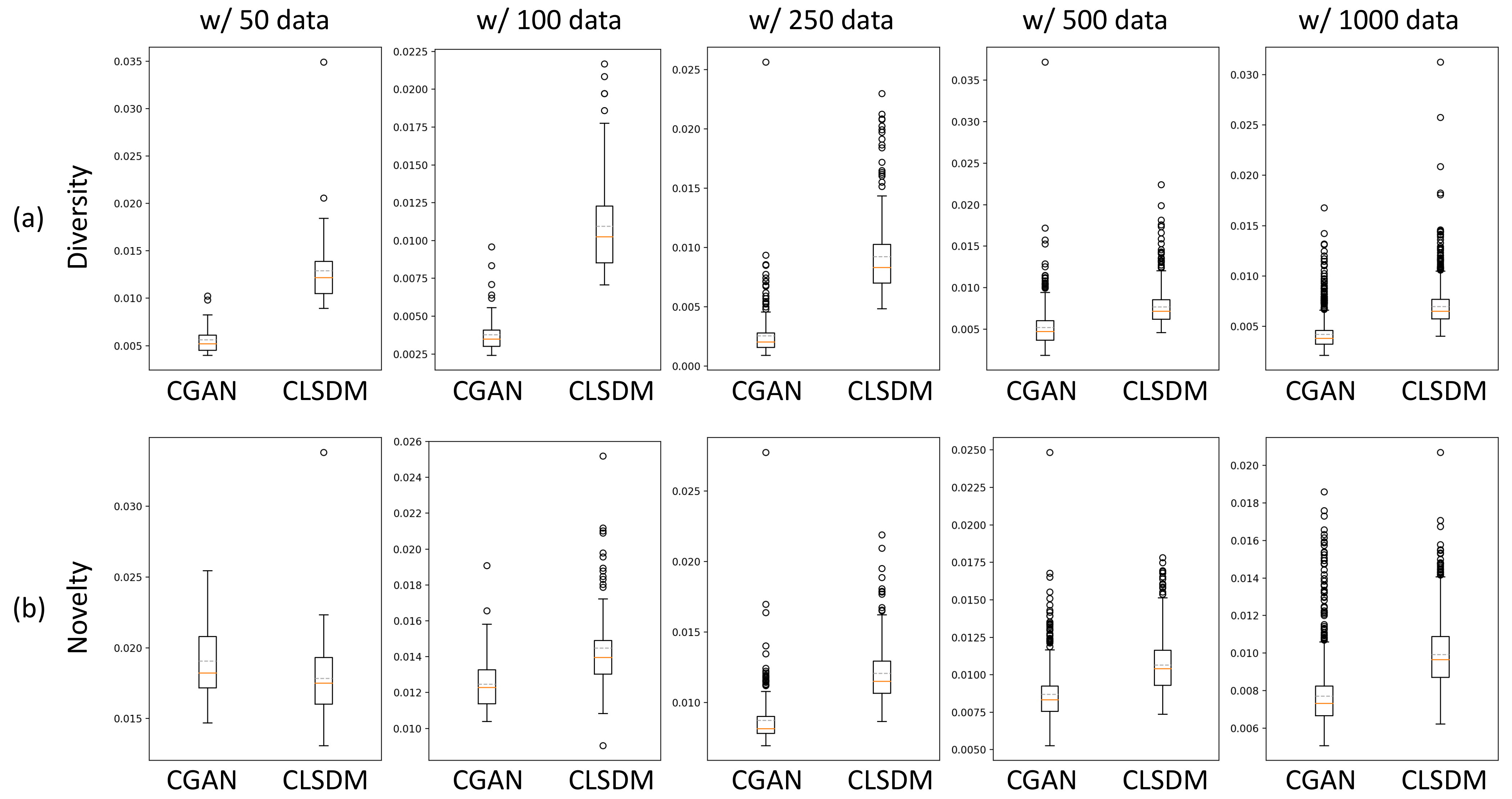}
    \end{center}
    \vspace{-2mm}
    \caption{
        \small Diversity ((a) $D_{intra}^{10}$) and novelty  ((b) $D_{inter}^{10}$) metrics for CGAN and CLSDM across dataset sizes.  
    }
    \label{fig:main_benchmark_conditional_intra_inter_dist}
\end{figure}

We also assess sampling exploration under conditioning to investigate whether the presence of a condition degrades diversity and novelty. By using the same $D^{10}_{intra}$ and $D^{10}_{inter}$ metrics, Fig.~\ref{fig:main_benchmark_conditional_intra_inter_dist} compares CLSDM and CGAN in terms of diversity and novelty of the conditioned samples. CLSDM generally maintains higher diversity and novelty than CGAN, similar to the unconditional case. Both models improve when given larger training sets, but CLSDM still shows better performance. Interestingly, CGAN’s novelty is higher at the lowest data setting with 50 samples. However, this is not due to CGAN exploring meaningfully, but rather because many CGAN outputs violate the area condition and effectively behave like unconstrained samples. These off-target shapes can appear “novel”, but they are not valid designs that satisfy the condition. In contrast, CLSDM’s samples adhere to the condition, so their novelty reflects the model's exploration within the constrained space.

In overall, \textit{DiffGeo}’s conditional diffusion CLSDM achieves a better balance between adherence to the condition and diversity of outputs. It supports the desired property with minimal sacrifice in variety, while the CGAN often either fails the condition or collapses the sampling diversity.

\revise{In conclusion, these benchmarking results demonstrate that \textit{DiffGeo} outperforms conventional GAN and variational approaches in data-scarce regimes. It achieves high-fidelity and diverse sampling with as few as 50 training shapes and enforces constraints more precisely. More importantly, \textit{DiffGeo} offers superior deployment flexibility by decoupling geometry generation from task-specific guidance, thereby eliminating the need for model retraining when design objectives change.}







\subsection{Task-Informed Data Generation for Surrogate-Based Optimization}
\label{sec:SBO}

After establishing \textit{DiffGeo}’s baseline performance, we next explore its integration into a surrogate-based optimization (SBO) workflow. In many aerodynamic design problems, engineers employ surrogate models for fast approximations of CFD simulations or experiments to optimize shapes. The training data for such surrogates is critical: it must capture the design space and relevant physics to guide the optimization effectively. Here, we investigate using \textit{DiffGeo} to sample training shapes more effectively that embed task-specific prior knowledge, and compare this approach with conventional design space exploration methods for data generation.

\subsubsection{Problem Setup}
We consider a 2D airfoil shape optimization task that maximizes the lift-to-drag ratio (L/D) of a NACA-0012 airfoil at Mach 0.85 and zero angle of attack in an inviscid flow. A geometric constraint is applied to maintain a maximum thickness of $12\%$ chord length. This setup represents a simplified transonic airfoil design task. The thickness constraint $C^I_{MT}$ is implemented by evaluating the airfoil’s thickness distribution $\cT(x)$ which is the vertical distance between upper and lower surface at each chord-wise position $x$, and building an inequality energy function for any violation of the $12\%$ chord limit. Specifically, we define:
\begin{equation}
    {C^I_{MT}} = \mathop {\max }\limits_{x \in (0,1)} \left( {\left|\left|\max (\cT(x) - 0.12 ,0)\right|\right|^2} \right)\;.
\end{equation}
$C^I_{MT}$ is nonzero only if the shape exceeds the thickness limit. 

In a typical SBO pipeline, there are two common ways to generate the surrogate’s training dataset: either (i) perturb an existing design via statistical random sampling with a low-dimensional parameterization, or (ii) collect designs from historical databases. We emulate both strategies as baselines:
\begin{itemize}
    \item \textbf{Baseline SM\#1 with Parametric Perturbations}: \revise{This strategy starts by fitting the NACA-0012 airfoil with a Class/Shape Transformation (CST) model to represent the upper and lower surfaces. Then new shapes are created by randomly sampling the CST coefficients within specified ranges using Latin Hypercube Sampling (LHS). We consider two variants of this strategy: a wide sampling range (SM\#1-1) using a 7th-order CST with 15 parameters, and a narrow sampling range (SM\#1-2) using an expertly tuned setting based on a reduced 3rd-order CST (8 parameters) \cite{Renganathan20}. For upper $\bw_u$ and lower $\bw_l$ surface coefficients, SM\#1-1 samples both coefficients between 0.1 and 2.5 times its baseline value. The trailing edge thickness $\delta_{TE}$ is varied in range of $[-10^{-3}, 10^{-3}]$. SM\#1-2 restricts perturbations to $\pm 30\%$ of the fitted NACA-0012 coefficients. This produces smoother but more constrained shape perturbations. Both SM\#1-1 and SM\#1-2 generate 500 sampled airfoils.}

    \item \textbf{Baseline SM\#2 with Historical Designs}: This approach follows the practice of using data available from previous projects. To emulate this idea, a small dataset of 50 profiles randomly selected from the UIUC airfoil database is built. These represent a mix of airfoil types, such as for aircraft wings, hydrofoils, etc., many of which are not tuned for transonic $L/D$ performance. This approach is easy to implement in practice but the data collected can be misaligned with the new design objective. 
\end{itemize}

\begin{figure}[tbh]
    \begin{center}
        \includegraphics[width=1\linewidth]{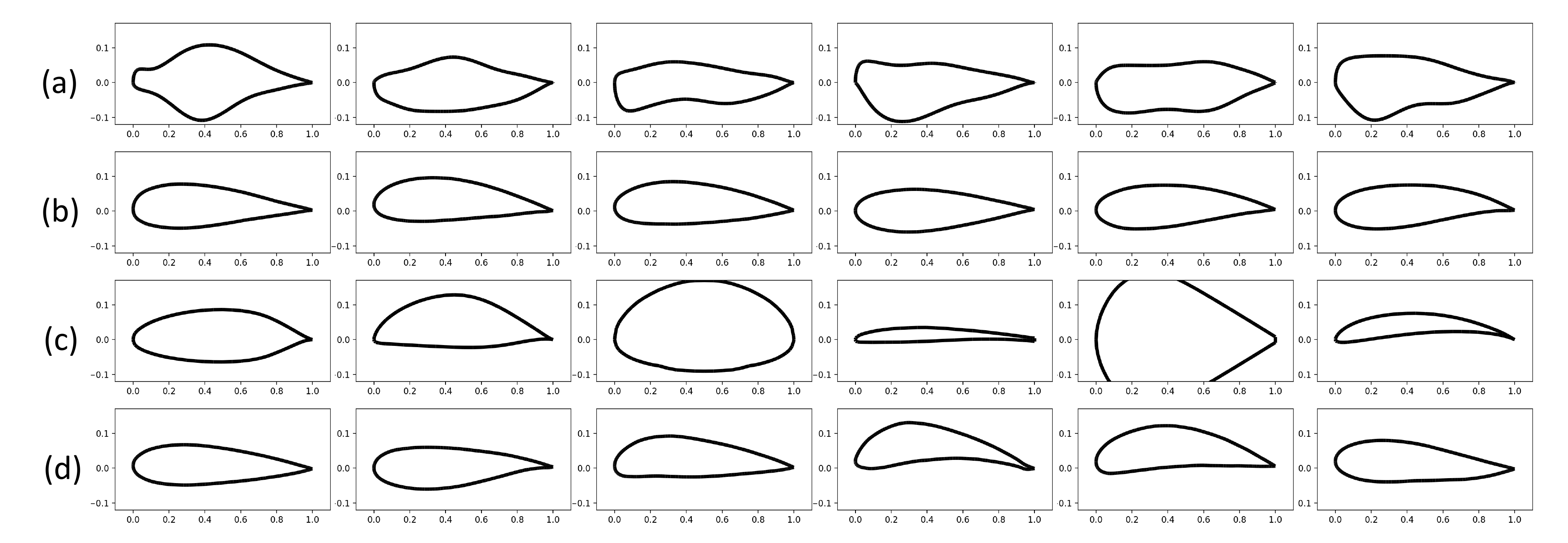}
    \end{center}
    \caption{
        \small \revise{Airfoil training sets: (a) SM\#1-1, (b) SM\#1-2, (c) SM\#2, (d) SM\#3.}
    }
    \label{fig:main_SBO_airfoils}
\end{figure}

\textit{DiffGeo} provides a third approach via its conditional sampling, denoted as SM\#3. \textit{DiffGeo} is trained on the same 50-profile dataset from SM\#2, thus starting with the same information as the historical data. Despite the dataset being small and not customized for this task, \textit{DiffGeo} can learn a latent geometry distribution from it. Then the conditional generation guided by the thickness constraint produces feasible and thickness-constrained shapes, using the CLSDM coupled with the energy function $C^I_{MT}$. By sampling 500 shapes through the sampling process described by Algorithm~\ref{alg:abs_sample_conditional_diffusion}, we obtain a training set for surrogate model that is task-informed while being diverse. Fig.~\ref{fig:main_SBO_airfoils} visualizes the each resulting airfoil datasets by SM\#1-1, SM\#1-2, SM\#2 and SM\#3.

\begin{figure}[!h]
    \begin{center}
        \includegraphics[width=1\linewidth]{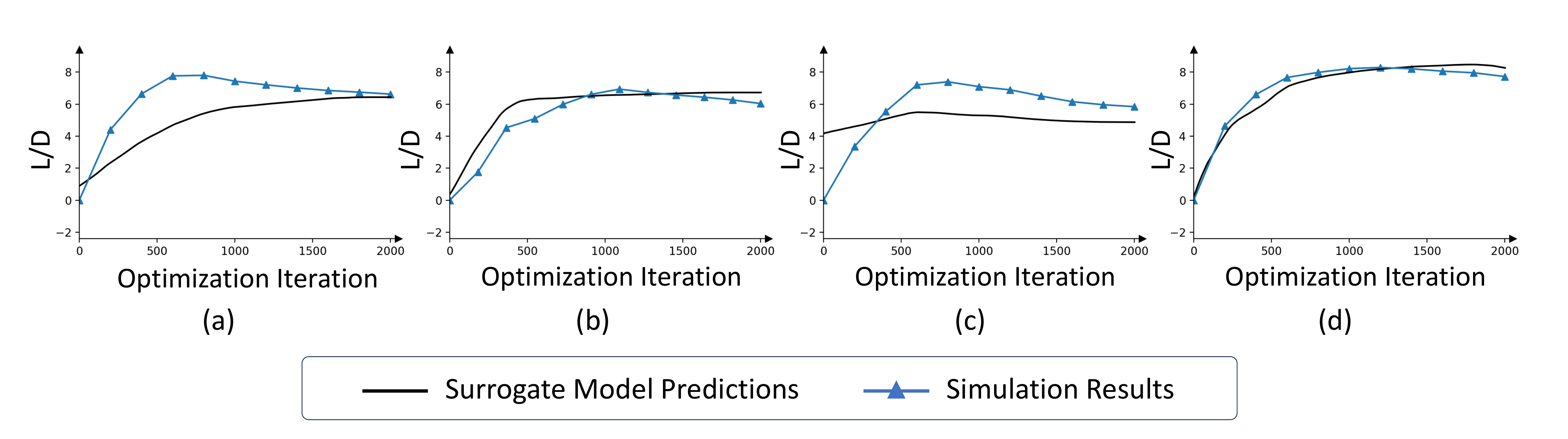}
    \end{center}
    \caption{
        \small \revise{Lift-to-drag ratio histories during surrogate-based optimizations with (a) SM\#1-1, (b) SM\#1-2, (c) SM\#2 and (d) SM\#3}.
    }
    \label{fig:main_SBO_optimized_history}
\end{figure}

\begin{figure}[!h]
    \begin{center}
        \includegraphics[width=1\linewidth]{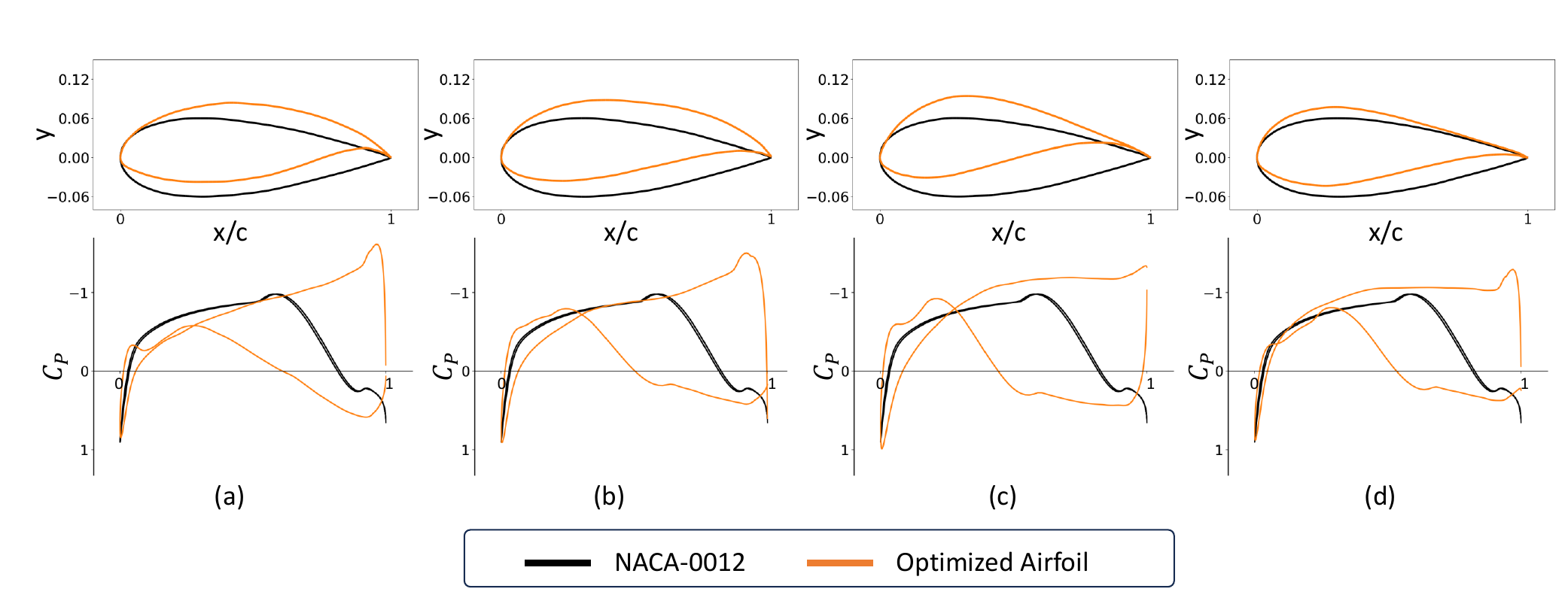}
    \end{center}
    \caption{
        \small \revisetwo{Shapes and $C_P$ distributions of NACA-0012 baseline and optimized airfoils by (a) SM\#1-1, (b) SM\#1-2, (c) SM\#2, and (d) SM\#3}.
    }
    \label{fig:main_SBO_optimized_airfoils}
\end{figure}

With these generated airfoils, we evaluate their aerodynamic performance using a triangulated CFD mesh built on NACA-0012\footnote{\scriptsize\url{https://github.com/su2code/Tutorials/blob/master/design/Inviscid_2D_Unconstrained_NACA0012/mesh_NACA0012_inv.su2}}. The \textit{DeepGeo} model ~\cite{Wei23b,Wei24a,Wei25a} is used to deform this mesh to each dataset profiles. Flow simulations are conducted by the Euler solver in \textit{SU2}~\cite{Economon16}, providing the lift $C_L$ and drag $C_D$ coefficients. The airfoils and corresponding aerodynamic performance are used to train surrogate models that predicts $C_L$ and $C_D$. The surrogate is a simple multi-layer perceptron (MLP) consisting of two hidden layers with ReLU activations. The input features are 120-dimensional vectors concatenated from the $x$ and $y$ coordinates of $60$ points sampled from the airfoil contour. We train each surrogate model for 500 epochs using Adam optimizer to minimize mean squared error (MSE) between predictions and simulated $C_L$ and $C_D$ in the training set.

The shape optimization uses LSM and its latent space for parameterization. The latent code $\bz$ is first initialized by encoding the baseline NACA-0012 airfoil using Equation~\ref{eq:agmin_f_z}. We then perform gradient-based optimization on $z$ to maximize L/D subject to geometric and regularization constraints. The optimization problem is:
\begin{equation}
    \bz^* = \argmin_\bz ||C_D(\bz)||^2 - ||C_L(\bz)||^2 + C^I_{MT}(V) + w_\bz||\bz||^2\;,
\end{equation}
where $\{C_D(\bz), C_L(\bz)\}$ are the lift and drag predicted by the surrogate for the shape corresponding to latent $z$. The term $w_z||z||^2$ with a small weight $w_z$ prevents the optimizer from being far from the latent manifold and penalizes extremely large latent perturbations. We use the Adam optimizer on $z$ with a learning rate of $10^{-4}$ and $2000$ iterations to ensure convergence.

\setcounter{table}{0}
\begin{table}[htbp]
  \centering
  \caption{Surrogate-based optimization results and surrogate model accuracy ($R^2$).}
    \begin{tabular}{lccc}
    \hline
    \textbf{Surrogate Model} & \textbf{Final Surrogate $L/D$} & \textbf{Final Simulation $L/D$} & \textbf{Surrogate Model Accuracy (R2 Score)} \\
    \hline
    SM\#1-1 & 6.4322  & 6.6236  & 0.3265  \\
    SM\#1-2 & \revise{6.7252}  & \revise{6.9370}  & \revise{0.7995}  \\
    SM\#2 & 4.8718  & 5.8405  & 0.2041  \\
    SM\#3 & \textbf{8.2603}  & \textbf{7.7188}  & \textbf{0.9620}  \\
    \hline
    \end{tabular}%
  \label{tab:main_SBO_results}%
\end{table}%

\subsubsection{Results}
Fig.~\ref{fig:main_SBO_optimized_history} shows the optimization histories with L/D versus iteration for each surrogate model. SM\#3, which relies on \textit{DiffGeo}, achieves the highest optimized L/D. Starting from the baseline L/D around 0.6, SM\#3's curve rises steadily to around 7.7 by 1200th iteration and plateaus. \revise{In contrast, SM\#1-1 improves to around 7.5 L/D but then drops. SM\#1-2 shows stable convergence but plateaus at a lower performance level (L/D=6.9). While the strict constraints in SM\#1-2 successfully prevent the model collapse observed in SM\#1-1, they also appear to overly restrict the design space, preventing the optimizer from reaching the higher-performance regions found by SM\#3. SM\#2 performs better than SM\#1-1 but underperforms compared to SM\#3. These results show that surrogate models trained on \textit{DiffGeo}’s conditional samples guide the optimization to an improved design, balancing shape validity with sufficient exploration.}

To quantify surrogate fidelity, Tab.~\ref{tab:main_SBO_results} reports the $R^2$ score of each surrogate model measured by comparing surrogate-predicted L/D to CFD-simulated L/D along the optimization path. SM\#3 reaches an $R^2$ of $0.96$, showing its near-perfect consistency with simulated performance. SM\#3's prediction can be trusted throughout the optimization. \revise{SM\#1-2 also performs well and achieves an $R^2$ of $0.80$. This is a significant improvement over the wide-range LSH sampling SM\#1-1 ($R^2=0.33$) and the random sampling SM\#2 ($R^2=0.20$), showing that expert tuning can produce functional surrogate models. However, SM\#3 still holds a distinct advantage. Since surrogate prediction is often the only indicator in real-world SBO applications, the superior reliability of SM\#3 and its tuning-free automation are critical for practical deployment.}

The final optimized shapes are compared in Fig.~\ref{fig:main_SBO_optimized_airfoils}. These four surrogate models produce distinctly different results. SM\#1-1, SM\#1-2 and SM\#2 generate airfoils with large cambers and aggressive deformations that lead to excessive drag and degraded actual performance. In contrast, SM\#3’s optimized airfoil is only a subtle modification of the baseline, yet achieves the highest L/D among all results. \revisetwo{The corresponding pressure-coefficient distributions in Fig.~\ref{fig:main_SBO_optimized_airfoils} further support this interpretation. Compared with the other sampling strategies, SM\#3 yields a smoother pressure recovery and avoids the pronounced leading-edge crossover behavior observed in some baseline cases, indicating that the surrogate trained with \textit{DiffGeo}-generated data provides a more consistent local design direction.} This result suggests that \textit{DiffGeo}’s samples helped the surrogate model generalize in the right region of design space around the baseline. \revisetwo{We emphasize that this SBO case study is intended to compare data-generation strategies for low-data surrogate-assisted optimization, rather than to establish a production-level transonic ASO benchmark. Since the surrogate is trained on a finite local design set around the NACA-0012 baseline and the flow evaluations use an inviscid Euler setup, the optimized shapes should be interpreted as workflow-validation results within this simplified local design region, rather than as aerodynamically certified final transonic designs.}

\revise{Note that SM\#3's advantage comes not only from generating diverse airfoils, but also from generating more effective ones. While both SM\#1 and SM\#3 used 500 samples, the random sampling in SM\#1 produced physically unrealistic geometries that offered limited benefit to the surrogate model. In contrast, \textit{DiffGeo} focused the sampling within a feasible domain through thickness conditioning. This highlights three key advantages of our framework: (1) Data Efficiency: \textit{DiffGeo} effectively produced useful designs from only 50 unrelated training shapes, outperforming naive random perturbations; (2) Deployment Flexibility: The model was repurposed for this new optimization goal without retraining, leveraging its geometry-performance disentanglement; and (3) Constraint Handling: Energy-based guidance targeted the feasible design space more effectively than conventional methods, which struggle to incorporate fine-grained constraints without manual filtering. Ultimately, integrating \textit{DiffGeo} into the SBO pipeline improved the surrogate model's accuracy ($R^2 \approx 0.96$) and guided the optimization to a superior design, demonstrating the practical value of generative sampling in engineering workflows.}

\subsection{3D Turbomachinery Blade Design Prototyping}
\label{sec:turbine}

Finally, we extend \textit{DiffGeo} to a challenging 3D design task: generating novel turbomachinery rotor blade shapes under minimal data conditions. This case evaluates \textit{DiffGeo}’s scalability to higher-dimensional geometries and its ability to handle more complex engineering constraints. We focus on a single-stage axial compressor rotor blade design problem, and assess if \textit{DiffGeo} can produce realistic, CFD-valid 3D blade designs from only a few existing examples, if it can explore nonlinear samples by learning from a linear data space, and if it can incorporate complicated constraints like spanwise thickness distributions or twist laws during generation.

\subsubsection{Data Collection}
Creating new rotor blade designs typically requires either a high-fidelity physics-based design cycle or large databases of existing blades, both of which are resource-intensive. To explore novel rotor blade designs from limited existing geometries, we simulate this setting by utilizing a small set of 75 blade geometries constructed by \citet{Fesquet24}. These 75 blades were generated by taking convex combinations of six baseline designs, forming a linear subspace spanned by the six reference blades. Specifically, 60 blades were obtained via farthest-point sampling in the convex hull of the six bases, and 15 blades were created by equal-proportion binary merges of two different base blades. This approach guarantees that all generated shapes are physically plausible, and avoid unrealistic or non-meshable shapes that random perturbations might cause (e.g., by randomly sampling in the compact and expressive 120-dimensional representation manually defined by Parablade~\cite{Agromayor21}). Fig.~\ref{fig:unconditional_blades}(a) shows a few examples of these interpolated blades. While they exhibit some variety, it is important to note that the dataset contains no fundamentally new shapes beyond the combinations of the six bases. According to the data processing inequality, deterministic interpolation cannot add and even lose information beyond the originals. We do not have access to the Parablade parametric representations of the six base blades. We only have these 75 interpolated geometries as point clouds. This makes the learning problem challenging: \textit{DiffGeo} must learn from a tiny and highly constrained dataset of 75 samples that essentially lie on a low-dimensional hyperplane in shape space.

The following investigations aim to assess whether \textit{DiffGeo} can (i) learn a valid generator from a limited set of interpolated blades, and (2) extend beyond this linear regime to generate feasible, nonlinear and controllable designs outside the convex hull.

\subsubsection{Unconditional Generation Quality}
\begin{figure}[!htb]
    \begin{center}
        \includegraphics[width=1\linewidth]{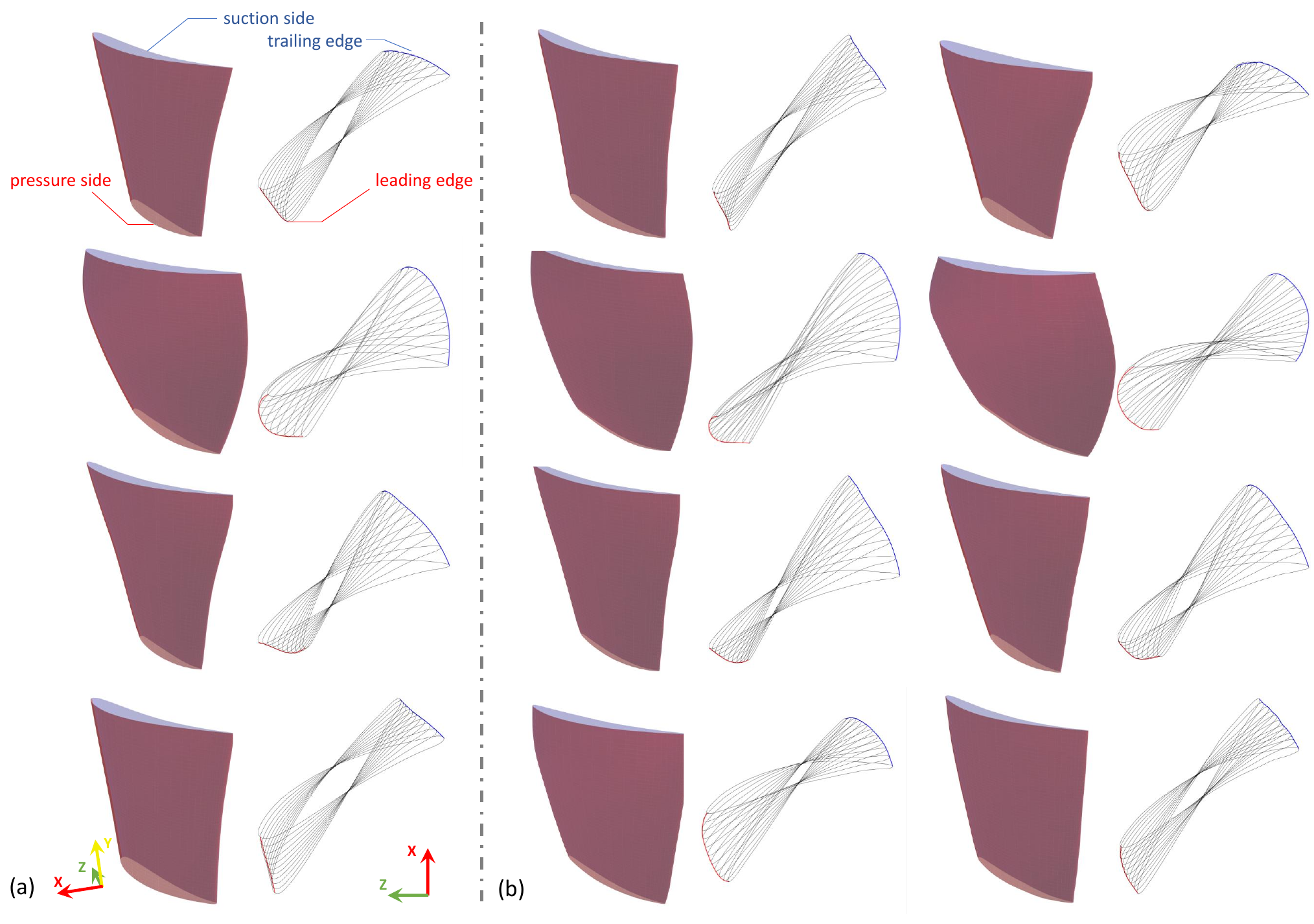}
    \end{center}
    \caption{
        \small 3D and 2D projected sectional views of (a) training dataset blades, (b) unconditionally generated blades by \textit{DiffGeo}.
    }
    \label{fig:unconditional_blades}
\end{figure}

The training of \textit{DiffGeo}'s LSM and LSDM for 3D blades follows the same procedure as in 2D. The template shape required by LSM is the vertex-wise average of all 75 blades, \revise{ and is discretized to $N=4000$ vertices}. Fig.~\ref{fig:unconditional_blades}(b) shows blades generated by unconditional \textit{DiffGeo}. They all exhibit valid shape properties: smooth surfaces, feasible spanwise twist, reasonable thickness and camber distributions, with no non-physical artifacts. We did not have to impose any explicit geometric filtering.

\begin{figure}[!htb]
    \begin{center}
        \includegraphics[width=1\linewidth]{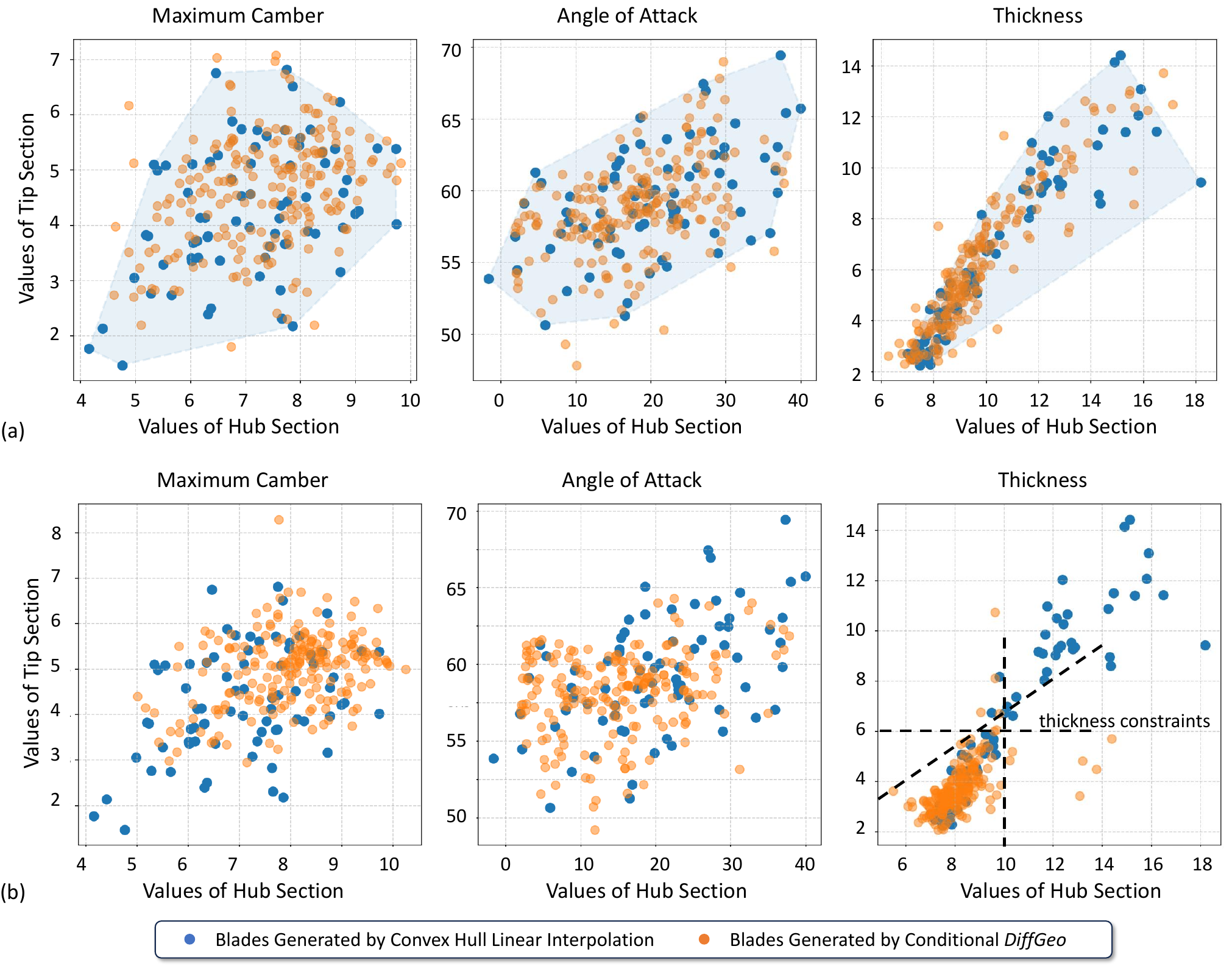}
    \end{center}
    \caption{
        \small Geometric statistics of blades by (a) unconditional and (b) conditional \textit{DiffGeo}.
    }
    \label{fig:blade_stats}
\end{figure}

\begin{figure}[!htb]
    \begin{center}
        \includegraphics[width=1\linewidth]{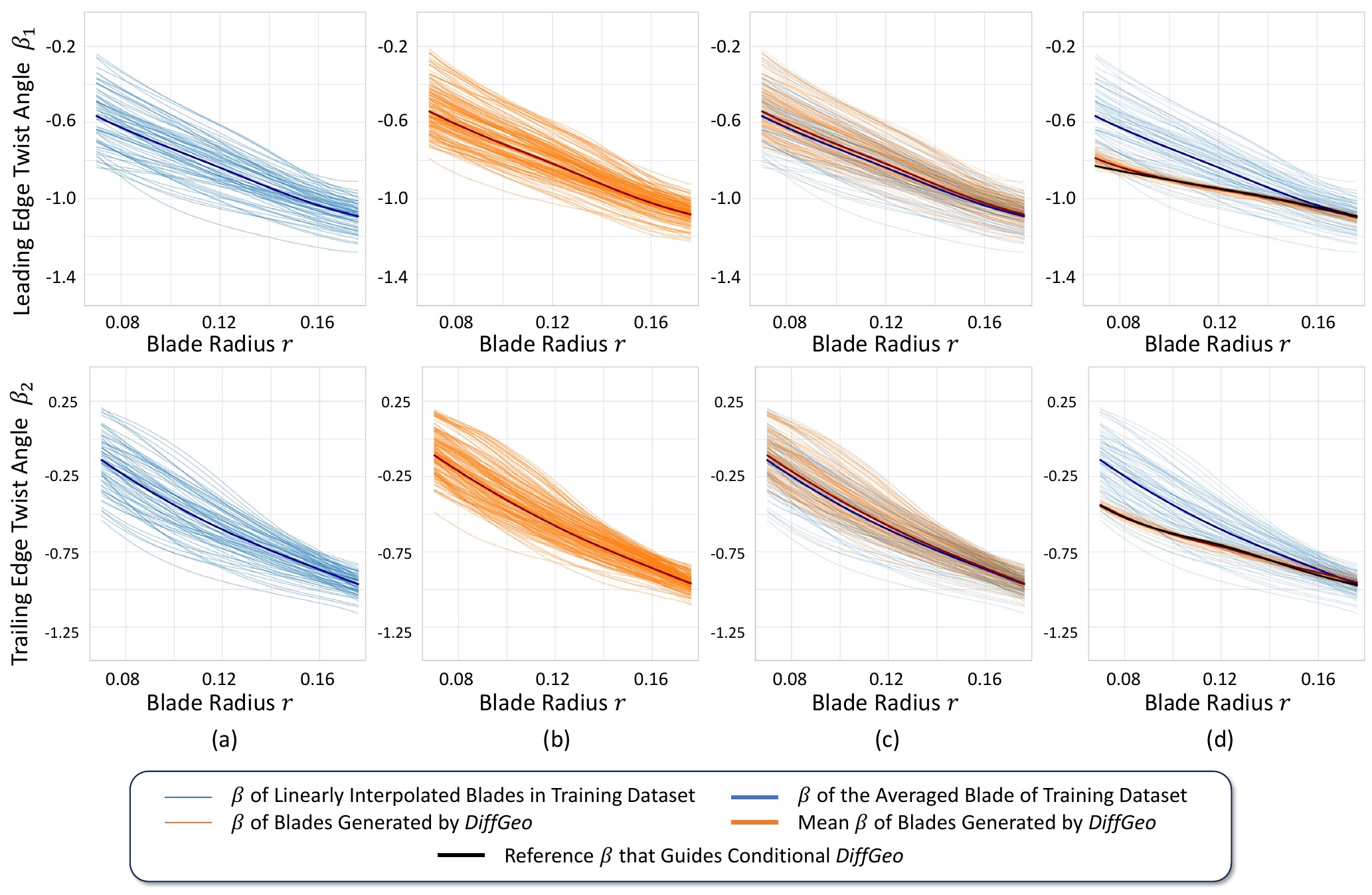}
    \end{center}
    \caption{
        \small Blade twist distributions: (a) training dataset, (b) unconditional \textit{DiffGeo}, (c) dataset vs. unconditional \textit{DiffGeo}, (d) dataset vs. twist-guided conditional \textit{DiffGeo}.
    }
    \label{fig:blade_beta}
\end{figure}

To ensure these qualitatively valid outputs also match the statistical distribution of the limited training data, several key geometric parameters between 200 \textit{DiffGeo}-generated blades and the 75 training blades are compared. In Fig.~\ref{fig:blade_stats}(a) plots several metrics for both sets, including the maximum thickness as a percentage of chord, maximum camber, and angles of attack all at the hub and tip sections. The generated blades (orange distribution) highly align with the training blades (blue distribution) for all these metrics. Similarly, Fig.~\ref{fig:blade_beta}(a-c) overlays the twist angle distributions (leading edge $\beta_1$ and trailing edge $\beta_2$ twist angles) of the training and generated blades. The individual $\beta$ profiles (in light lines) vary, but the averaged twist law of \textit{DiffGeo}'s generation (in orange solid lines)  matches the $\beta$ of mean training set blade (in blue solid lines). In summary, \textit{DiffGeo} matches the in-convex-hull distribution generated by linear combinations of the six base blades.

\begin{figure}[!htb]
    \begin{center}
        \includegraphics[width=1\linewidth]{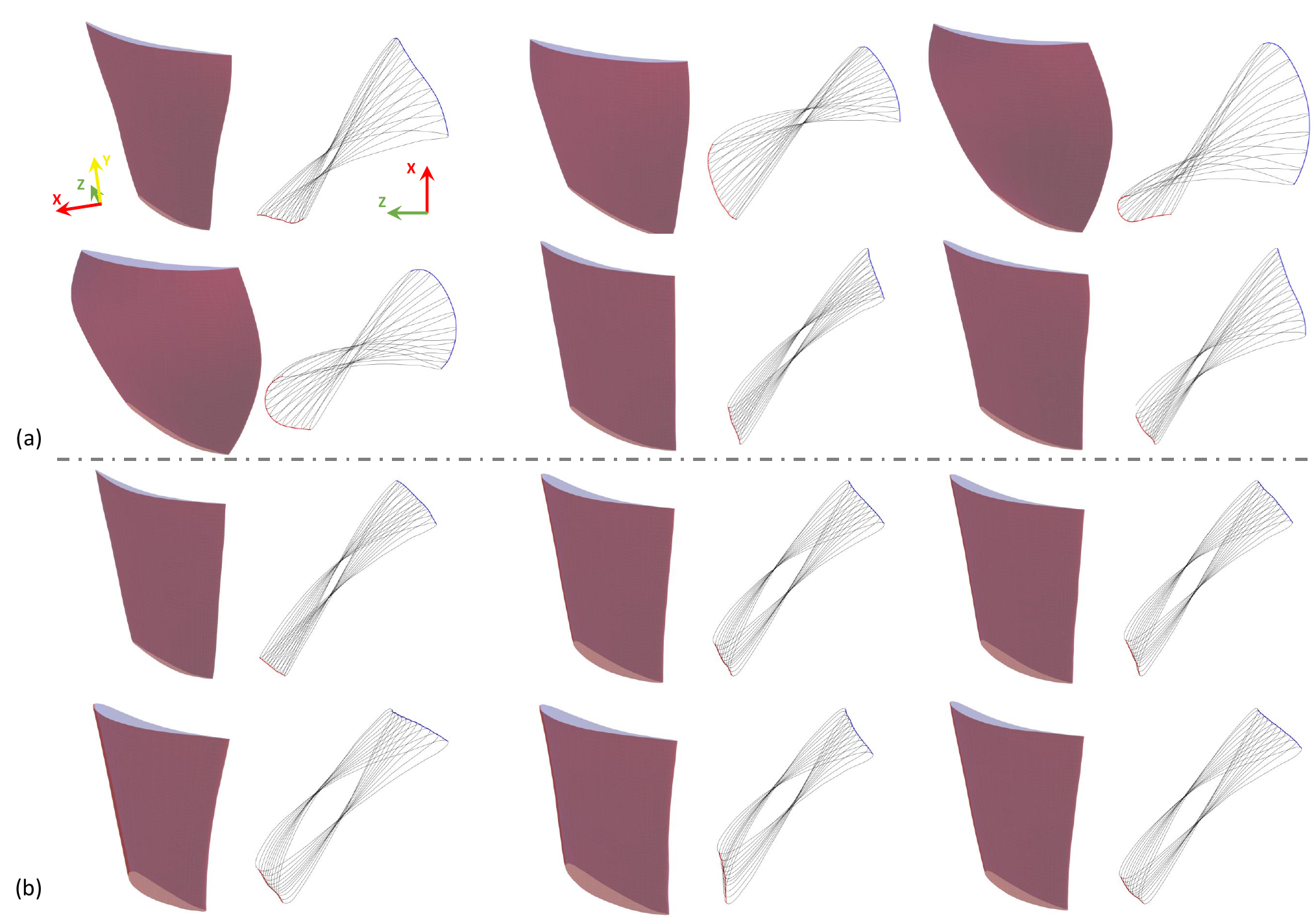}
    \end{center}
    \caption{
        \small 3D and 2D projected sectional views of conditional \textit{DiffGeo} blades guided by (a) thickness constraints, (b) reference twist law.
    }
    \label{fig:conditional_blades}
\end{figure}

\subsubsection{Controllability of Guided Sampling}
In this section, we investigate \textit{DiffGeo}'s controllability on a 3D task by imposing multiple simultaneous constraints. Specifically, we guide \textit{DiffGeo} to generate blades that satisfy certain thickness requirements at both hub and tip. This simulates an engineering scenario where the blade must meet thickness criteria at different span locations for structural or performance considerations. Three inequality constraints are imposed in the normalized scale of dataset: (i) the hub section’s maximum thickness must be less than 10, (ii) the tip section’s max thickness less than 6, and (iii) the ratio of tip thickness to hub thickness less than $2/3$. These constraints are higher-dimensional and coupled, and are challenging to handle with traditional design of experiments or GAN/VAEs without specialized training. However, they reflect more realistic design rules in turbomachinery.

To integrate these requirements within \textit{DiffGeo}, let $\cT_{hub}(x)$ and $\cT_{tip}(x)$ denote the thickness distributions along the chord for the hub and tip cross-sections of a blade, similar to 2D case. Let:
\begin{align}
    C^I_{hub} &= \max (\max_x \cT_{hub}(x)-10, 0) \;, \\
    C^I_{tip} &= \max (\max_x \cT_{tip}(x)-6, 0) \;, \\
    C^I_{t2h} &= \max (\frac{\underset{x}\max \cT_{tip}(x)}{\underset{x}\max \cT_{hub}(x)}- \frac{2}{3}, 0) \;,
\end{align}
which we sum into a simple composite inequality constraint function $C^I = C^I_{hub} + C^I_{tip} + C^I_{t2h}$. Blades are then generated following the guided sampling steps described in Algorithm~\ref{alg:abs_sample_conditional_diffusion}, where $C^I$ is added as in Equation~\ref{eq:energy_guidance} to push the generated shape toward satisfying all three thickness limits simultaneously.

Some qualitative results are shown in Fig.~\ref{fig:conditional_blades}(a). The blade geometries remain smooth and continuous, and patterns such as camber and twist distribution are still diverse, showing valid properties. They exhibit visibly thinner profiles at the tip and/or hub compared to unconditional generated samples. Fig.~\ref{fig:blade_stats}(b) quantitatively demonstrates the effect of thickness constraints for 200 guided samples (in orange points), where the thickness metrics have clearly shifted while the other metrics, including camber and angle of attacks ($AoA$), remain aligned with the training data (in blue points). \textit{DiffGeo} successfully isolates the effect of enforcing thickness constraints as it generates new shapes that fulfill the constraints without distorting unrelated aspects of the blade geometry. This highlights \textit{DiffGeo}’s ability to handle high-dimensional constraints, achieving results that would be intractable for brute-force sampler or retrained models under limited data.

\subsubsection{Nonlinear Exploration Beyond Linear Interpolation}
A central question for this case study is whether \textit{DiffGeo} can produce truly novel blades that are not simply interpolations of the limited training set. Since the training blades lie in the convex hull of six existing designs, any design outside that convex hull represents a nonlinear combination or extrapolation that introduces new geometry. 

To answer this question quantitatively, \revise{we perform a linear reconstruction error analysis to calculate the convex-hull residual}. Let the six base blades be $\{S_1,...,S_6\}$, the interpolated training data be $\{\bar{S}_1,...,\bar{S}_{75}\}$ and $K$ generated blades be $\{M_1,...,M_K\}$. The reconstruction errors of training dataset $\mathcal{E}_{data}$ and of \textit{DiffGeo}'s generation $\mathcal{E}_{gen}$ are defined as the solutions to the following optimization problems:
\begin{align}
    \mathcal{E}_{data}^{(i)} & =\underset{\{w_1^{(i)},...,w_6^{(i)}\}} {\mathrm{min}} \frac{1}{75}\sum_{j=1}^{75} \bigl\| \bar{S}_j-\sum_{k=1}^6 w_k^{(i)}S_k \bigr\|^2 \;,\\
    \mathcal{E}_{gen}^{(i)} & =\underset{\{w_1^{(i)},...,w_6^{(i)}\}} {\mathrm{min}} \frac{1}{K}\sum_{j=1}^{K} \bigl\| M_j-\sum_{k=1}^6 w_k^{(i)}S_k \bigr\|^2 \;,
\end{align}
where $\{w_1^{(i)}, \dots, w_6^{(i)}\}$ are nonnegative interpolation weights for the $i$-th tested blade subject to $\sum_{j=1}^6 w_j = 1$, ensuring each reconstruction lies within the convex hull of the base blades. \revise{The convex hull defines the bounds of what can be achieved through simple linear interpolation of the baseline data. Therefore, a non-zero residual is a highly positive indicator in generative design: It shows that \textit{DiffGeo} is not simply mixing existing data points and memorizing training data. Instead, it has successfully learned the underlying non-linear manifold. A higher residual indicates a greater deviation from the linear subspace, reflecting the model's capability for non-linear extrapolation to discover truly novel designs.}

\begin{figure}[!htb]
    \begin{center}
        \includegraphics[width=1\linewidth]{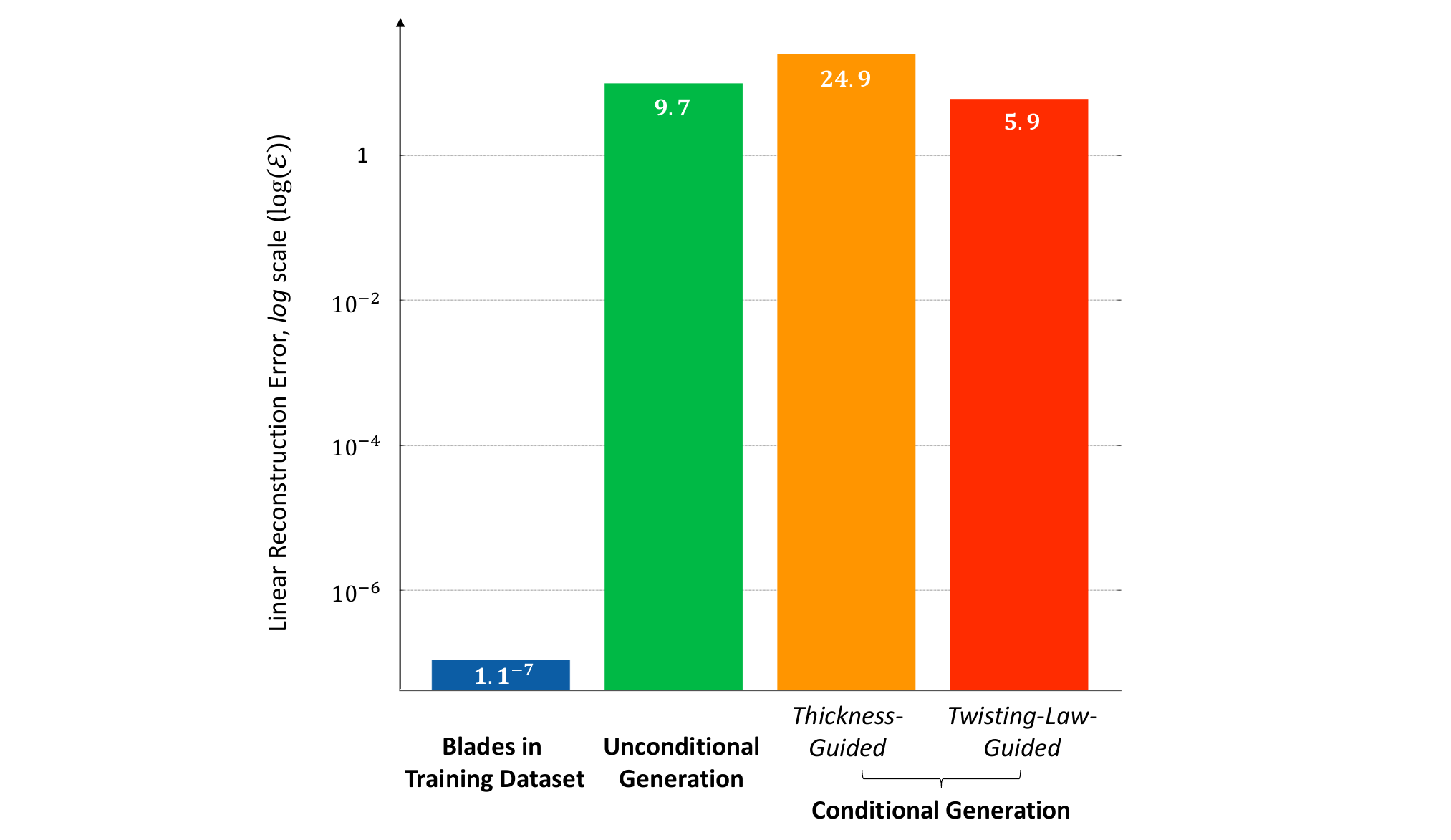}
    \end{center}
    \caption{
        \small Linear reconstruction errors of interpolated dataset blades, \textit{DiffGeo}'s unconditional, thickness-guided and twisting-law-guided conditional outputs.
    }
    \label{fig:linearity_check}
\end{figure}

By definition, the training blades have near-zero reconstruction error, which is confirmed in Fig.~\ref{fig:linearity_check}.
In comparison, \textit{DiffGeo}-generated blades show substantially larger convex-hull residuals: 9.7 for unconditional generation and 24.9 for thickness-guided generation, measured as the average per-surface-point L2 distance. For reference, the typical hub-section chord length is about 80, meaning these residuals correspond to roughly $12\%$ and $31\%$ of the hub chord, respectively. Such magnitudes are well beyond numerical tolerance, indicating that the generated blades cannot be represented by any convex combination of the six bases, and that \textit{DiffGeo} explores geometry outside the range accessible to linear interpolation.

To further evaluate nonlinear conditional controllability, we compare \textit{DiffGeo} with the linear convex interpolation under a new set of design constraints. Specifically, we define four  specifications: hub maximum thickness of 12, tip maximum thickness of 8, hub angle of attacks of $10^{\circ}$ and tip angle of attacks of $55^{\circ}$. As can be observed in Fig.~\ref{fig:blade_stats}, each specification lies well within the convex hull spanned by the six base blades. For \textit{DiffGeo}, we apply energy-based guidance and generate 20 guided samples. For convex interpolation, we optimize nonnegative weights with different random initialization over the six bases to minimize the same energy.

Despite each of the target being in-convex-hull, the linear interpolation struggles to satisfy all constraints simultaneously. The hub AoA incurs a large residual, indicating a strong trade-off between these targets. In contrast, \textit{DiffGeo} achieves balanced and low total error across all specifications. Tab.~\ref{tab:blade_diffgeo_linear_comparison} shows the quantitative results. In terms of generation diversity, linear interpolation's mean standard deviation across surface points is $0.31$, while \textit{DiffGeo} produces a larger mean standard deviation of $1.40$ under the same target. This comparison clearly demonstrates that \textit{DiffGeo}’s nonlinear latent space offers additional degrees of freedom for DSE than linear convex interpolation, while maintaining geometric validity and without sacrificing sample diversity.

\begin{table}[htbp]
  \centering
  \caption{Constraint errors of conditional generations under in-convex-hull target using \textit{DiffGeo} and linear convex interpolation.}
    \begin{tabular}{lccccc}
    \hline
    DSE Methods & \multicolumn{1}{l}{Error of $\cT_{hub}$} & \multicolumn{1}{l}{Error of $\cT_{tip}$} & \multicolumn{1}{l}{Error of ${AoA}_{hub}$} & \multicolumn{1}{l}{Error of ${AoA}_{tip}$} & \multicolumn{1}{l}{Total Error}\\
    \hline
    \textit{DiffGeo} & 1.037  & 0.573  & 0.448  & 0.629  & 2.687\\
    Linear Convex Hull & 0.018  & 0.578  & 6.353  & 0.859  & 7.808\\
    \hline
    \end{tabular}%
  \label{tab:blade_diffgeo_linear_comparison}%
\end{table}%

\subsubsection{Integrating \textit{DiffGeo} with Mean-Line Design}
Finally, we show how \textit{DiffGeo} can interface with a conventional mean-line design tool to streamline 3D blade prototyping. The mean-line design offers a fast and analytical means of exploring preliminary performance. A typical design workflow for a new rotor blade starts with global specifications, including expected mass flow rate, pressure ratio and rotational speed. Then the mean-line design tool performs one-dimensional through-flow analysis to solve conservation equations along a representative stream surface between hub and shroud, which produces spanwise distributions of key aerodynamic quantities, including flow angles, blade loading and annulus areas. A key output is the spanwise twist law $\beta$, which prescribes the blade metal angle from hub to tip, and ensures proper incidence and work distribution across the rotor. Along with the chord and solidity distributions, $\beta$ defines a family of quasi-3D sections from which the blade geometry can be reconstructed. Designers then create a 3D blade geometry that realizes this twist and meets other loading criteria, often by manually adjusting and stacking 2D airfoil sections which is a time-consuming process and a bottleneck for rapid conceptual design exploration. We propose using \textit{DiffGeo} to automate the generation of candidate 3D blade geometries that satisfy the mean-line prescribed twist, thus providing a transformed workflow: instead of manually constructing 3D blades, \textit{DiffGeo} proposes viable candidates to choose from as ready starting points for subsequent CFD-based analysis and optimization. This transition can reduce design complexity and workloads.

In this investigation, we seek to design a single-stage axial rotor at 0.3 mach and sea-level conditions. The design specifications include pressure ratio around 1.59 and rotation speed at 1800 rad/s. The reference twist laws $\hat{\beta}_1$ and $\hat{\beta}_2$ are obtained from an interactive mean-line design software that parameterizes the twist distribution with Bezier curves. The reference $\hat{\beta}_1$ and $\hat{\beta}_2$ are incorporated as guidance in \textit{DiffGeo}'s generation by formulating an energy term $C^E=C_{\beta_1}^E+C_{\beta_2}^E$, where:
\begin{align}
    C_{\beta_1}^E &= \int_{r_h}^{r_t} \bigl\|\hat{\beta}_1(r) - \beta_1(r,M)\bigr\|^2 dr \;, \nonumber \\
    C_{\beta_2}^E &= \int_{r_h}^{r_t} \bigl\|\hat{\beta}_2(r) - \beta_2(r,M)\bigr\|^2 dr \;,
    \label{eq:beta_constraint}
\end{align}
with $r_h$ and $r_t$ being the hub and tip radii, and $\beta_1(r,M)$ and $\beta_2(r,M)$ the corresponding twist distributions of a generated blade $M$. In practice, we discretize these integrals into sums over 50 spanwise sections. Minimizing $C_E$ during generation will encourage the blade’s twist to follow the mean-line solution.

Because matching an entire twist distribution is a stringent constraint, we apply enhanced conditional sampling as described in Algorithm~\ref{alg:abs_sample_enhanced_conditional_diffusion} with $N^T=3$ and $T^*=650$ to improve convergence. 300 blades targeted at the reference twist laws are generated, and Fig.~\ref{fig:conditional_blades}(b) shows some of them. Through the projected view of stacked spanwise sections, these blades have the preferred characteristic twist and still exhibit variability in other aspects like camber and thickness distribution. The $\beta$ distributions of conditional blades are compared in Fig.~\ref{fig:blade_beta}(d) (in orange lines) with the dataset distribution (in blue lines) and the reference target (in black solid lines). The guided blades indeed closely follow the reference twist law on average, showing that the energy guidance has successfully induced \textit{DiffGeo} to meet the mean-line design output.

\begin{figure}[!tb]
    \begin{center}
        \includegraphics[width=1\linewidth]{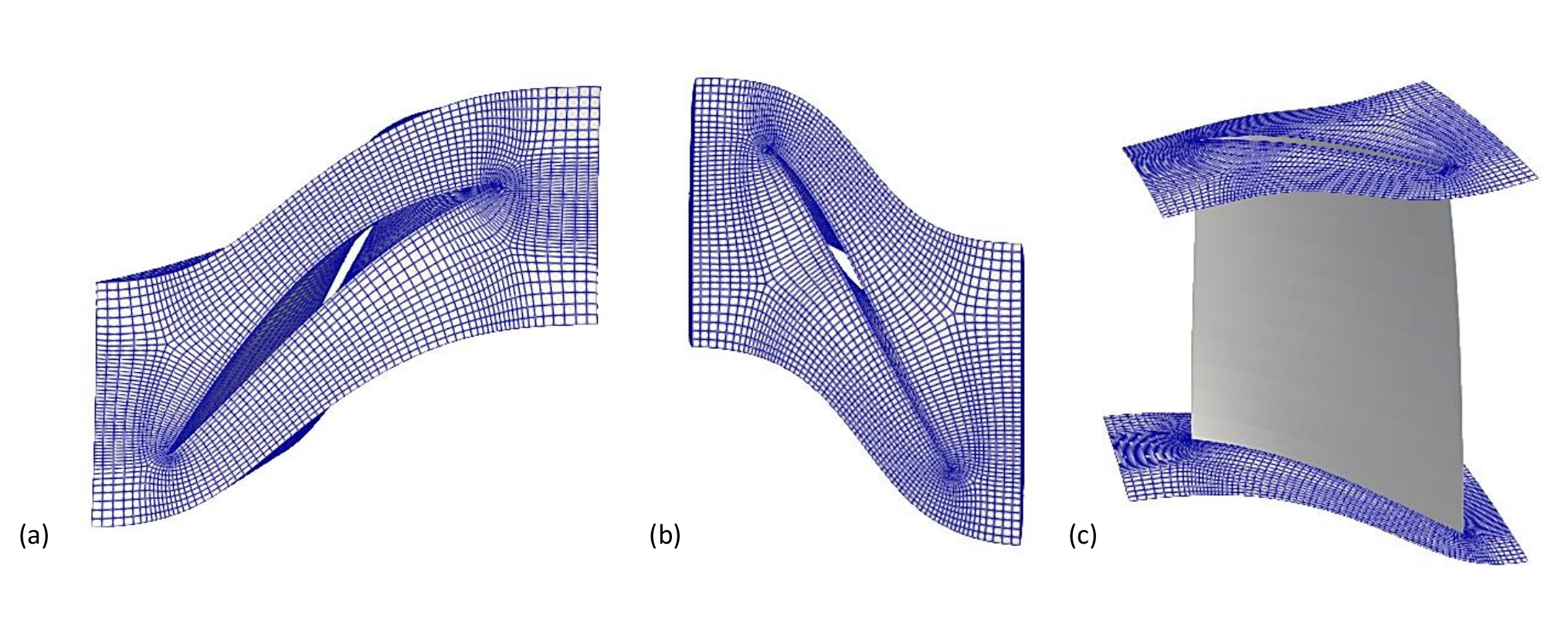}
    \end{center}
    \caption{
        \small CFD mesh for efficiency validation: (a) hub view, (b) tip view, (c) side view.
    }
    \label{fig:blade_simulation_mesh}
\end{figure}

\begin{figure}[!htb]
    \begin{center}
        \includegraphics[width=1\linewidth]{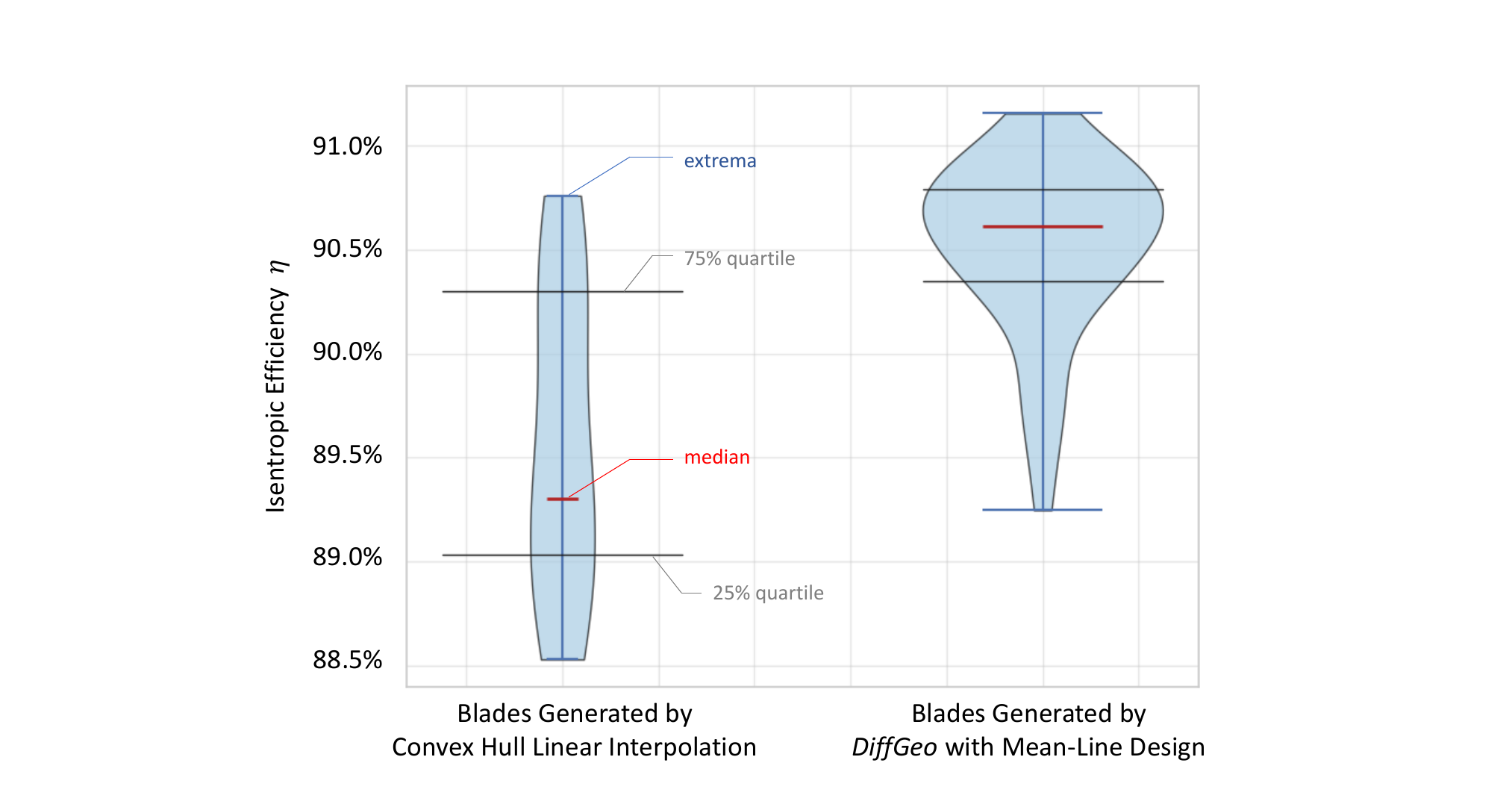}
    \end{center}
    \caption{
        \small Isentropic efficiency $\eta$ distributions for linearly convex-interpolated and conditional \textit{DiffGeo} blades.
    }
    \label{fig:isentropic_efficiency}
\end{figure}

One side effect of the strong twist guidance is that a fraction of the sampling can produce invalid shapes, accounting for $26\%$ of all samples. These invalid blades have less smooth surfaces when the latent code was pushed too far. However, this is not prohibitive since those invalid cases are easily filtered out with visual check or simple geometric validity check. As a result, 222 valid blades were left that meet the twist requirement. This $74\%$ success rate is still a favorable trade-off, given that producing a manually twist-aligned blade normally requires significant labor.

We then evaluate the aerodynamic performance of these blades through CFD simulations. To evaluate all blade profiles within a reasonable amount of time, we use a relatively lower-fidelity setup \footnote{\url{https://dafoam.github.io/mydoc_tutorials_aero_rotor37.html}} with DAFoam's RANS solver~\cite{He20d}. Each profile is meshed by deforming from a template 40k-cell CFD mesh originally built on Rotor 37 that only models the blade passage (shown in Fig.~\ref{fig:blade_simulation_mesh}), by using the \textit{DeepGeo}~\cite{Wei24a,Wei25a} model. \revise{To reliably handle the wide range of generated sectional variations, this deformation pipeline is augmented with explicit geometric constraints that enforce planar inlet/outlet boundaries and exact rotational consistency across the periodic patches.} The simulation produces shaft torque $C_{mz}$, pressure ratio $PR$, single-channel mass flow rate $\dot m$, from which we compute the isentropic efficiency $\eta$ defined as a ratio between the rotor's ideal work $W_{ideal}$ and actual work $W_{actual}$:
\begin{align}
    \eta &= \frac{W_{ideal}}{W_{actual}} \nonumber \\
         &= \frac{c_p\bigl(T_{0,out,ideal}-T_{0,in}\bigr)}{\frac{\omega \cdot C_{mz}}{\dot m}} \nonumber \\
         &= \frac{c_p \cdot T_{0,in} \cdot \bigl(\frac{p_{0,out}}{p_{0,in}}^\frac{\gamma-1}{\gamma}-1\bigr)}{\frac{\omega \cdot C_{mz}}{\dot m}}\\
         &= \frac{c_p \cdot T_{0,in} \cdot \bigl({PR}^\frac{\gamma-1}{\gamma}-1\bigr)}{\frac{\omega \cdot C_{mz}}{\dot m}}\;, \nonumber
\end{align}
where $\omega$ is the rotor’s rotation speed, $c_p$ the specific heat and $\gamma$ the specific heat ratio. $T_{0,in}$ and $T_{0,out,ideal}$ are the inlet total temperature and outlet ideal temperature, and $p_{0,in}$, $p_{0,out}$ are the corresponding total pressures. For comparison, we perform the same evaluation for the 75 convex hull interpolation blades in the training set. During this process, 193 generated blades and 55 dataset blades produce converged simulation results. The design candidates are selected if the blade's total pressure ratio close to the initial specifications (i.e. $1.6\geq {PR} \geq 1.58$), resulting in a selection of 45 generated blades ($20.3\%$ of valid generations) and 10 dataset blades (13.3\% of valid dataset profiles).

\begin{table}[htbp]
  \centering
  \caption{Sampling efficiency comparison between convex hull interpolation and \textit{DiffGeo} conditional sampling.}
    \begin{tabular}{lcc}
    \hline
    DSE Methods & Convex Hull Interpolation & DiffGeo + Mean Line Design \\
    \hline
    Number of Total Samples & 75     & 300 \\
    \hline
    Number and Ratio of  &        &   \\
    $\quad$ Valid Samples & 75 (100\%) & 222 (74.0\%) \\
    $\quad$ Successful Simulations & 55 (73.3\%) & 193 (86.9\%) \\
    $\quad$ Design Candidates & 10 (13.3\%) & 45 (20.3\%) \\
    \hline
    Isentropic Efficiency Median & 89.3\% & 90.7\% \\
    \hline
    \end{tabular}%
  \label{tab:turbine_sample}%
\end{table}%

Fig.~\ref{fig:isentropic_efficiency} summarizes the efficiency results. \textit{DiffGeo}-generated blades (on the right hand side) show consistently higher efficiency than the interpolated ones (on the left hand side). The median $\eta$ of the generated blades is about 90.7\%, whereas the convex hull interpolations’ median is 89.3\%. Additionally, the top end of the \textit{DiffGeo} efficiency range exceeds that of any interpolated blade. Statistics in Table~\ref{tab:turbine_sample} show that \textit{DiffGeo}’s conditional sampling guided by mean-line design outputs produces more valid designs, more high-performance candidates, and a higher median efficiency than the linear interpolation approach, even it starts from the same six baseline designs. We note that this comparison is based on a limited number of selected design candidates. The reported efficiency gains should be interpreted as indicative.

We also applied the linear convex-hull reconstruction test to the \textit{DiffGeo}-generated blades. The mean reconstruction error measured as averaged per-surface-point L2 distance is 5.9, indicating that the high-efficiency designs lie outside the convex hull of the bases (see Fig.~\ref{fig:linearity_check}). Their performance gains arise from nonlinear geometric variations in \textit{DiffGeo}’s latent space rather than from simple linear combinations of the base blades.

\revise{This case study confirms three critical capabilities of \textit{DiffGeo} in 3D: (1) Data Efficiency: Despite being trained on a limited linear subspace defined by only six base profiles, the model successfully learned a valid 3D design manifold; (2) Flexibility: The same pre-trained model was adapted to enforce disparate constraints without retraining; and (3) Nonlinear Extrapolation: \textit{DiffGeo} generated novel, high-efficiency designs that lie outside the training data's convex hull, effectively discovering superior geometries that were inaccessible to the baseline interpolation method.}

\section{Conclusion}
\label{sec:conclusion}
\revise{This work presents \textit{DiffGeo}, a latent space diffusion-based generative framework for data-efficient and controllable design space exploration. By reframing the generative process as a reusable shape sampler, we demonstrate that \textit{DiffGeo} serves as a fundamental tool for real-world design workflows. Through extensive experiments in both 2D airfoil and 3D turbomachinery blade cases, we validated three core capabilities of our approach: (1) high data efficiency, enabling high-fidelity generation with as few as 50 training airfoil samples and 6 blade samples; (2) task-agnostic adaptability, allowing the same pre-trained model to be repurposed for varying objectives without retraining; and (3) controllability under high-dimensional constraints, achieved via energy-based guidance for precise adherence to complex engineering requirements.}

From a broader perspective, \textit{DiffGeo} introduces a new paradigm for early-stage design exploration. Rather than relying on handcrafted parameterizations or trial-and-error sampling, designers can now directly generate feasible and diverse candidates that comply with desired performance constraints all in a one-shot and differentiable fashion. This capability greatly accelerates the conceptual design stage and enhances the quality of candidate designs available for downstream optimization. Our SBO experiments on 2D airfoils shows that this guided generative sampling strategy can be used to improve surrogate model's accuracy and final optimized performance, compared to baseline strategies based on random statistical sampling or historical data. Our 3D blade case study further confirms that \textit{DiffGeo} not only produces valid shapes outside the original linear interpolation space, but also leads to designs with higher aerodynamic efficiency.

We also acknowledge the current boundaries of the framework. \textit{DiffGeo}'s exploration is implicitly bounded by the training data manifold, making it an effective tool for interpolation and variation rather than conceptual invention of new topologies. Furthermore, a trade-off exists between strict constraint adherence and geometric smoothness under strong energy-based guidance. However, these limitations characterize the tool’s specific role in refining and exploring existing design concepts, and they provide a clear baseline for future advancements in manifold-constrained generation.

While this work focuses on geometric constraints in aerodynamic shape design, the core framework of \textit{DiffGeo} is not limited to geometry alone. The latent diffusion sampler and energy-based conditioning mechanism form a general interface for embedding design principles into generative exploration. Although this paper does not yet incorporate physical design objectives, the structure of our method inherently accommodates extensions where guidance is informed by physics--such as aerodynamic performance, structural stress or thermal constraints--through differentiable approximations or adjoint solvers. More importantly, the theory is not constrained to single physical domain. In the future, \textit{DiffGeo} has the potential to support multidisciplinary design scenarios, where diverse engineering requirements can be jointly encoded and explored in a unified generative framework.

We believe this work paves the way for democratizing generative design. The ability to train models on small datasets, incorporate expert intuition as energy functions, and generate high-performing designs without intensive tuning makes \textit{DiffGeo} accessible to teams with limited data or ML infrastructure. This lowers the entry barrier to AI-driven design and enables non-specialist engineers to explore new ideas with greater freedom and confidence.

For future studies, we see several opportunities to extend this work. First, we plan to incorporate physics-informed guidance into the sampling process by integrating adjoint-based solvers, enabling the generative model to address performance objectives such as lift-to-drag ratio, stress distribution or thermal compliance. To make this feasible in practice, we also aim to develop more efficient conditional sampling algorithms that reduce the number of energy function evaluations, thereby limiting the computational cost and solver calls during sampling. Second, we intend to enhance the robustness of the framework, including the study of sampling stability when operating near the edges of the learned latent manifold of designs, and introducing uncertainty-aware strategies to quantify generation confidence or constraint satisfaction likelihood. Finally, we plan to embed \textit{DiffGeo} into an interactive design workflow, where human engineers collaborate with the generative model in the conceptual design stage to provide online constraints, interpret generated results and enable exploration in real time.

In summary, \textit{DiffGeo} establishes a new approach to data-driven design space exploration. Its demonstrated performance in both 2D and 3D cases shows that it can generate valid, novel and constraint-satisfying shapes under severe data limitations. We hope this work will inspire further research in controllable generative modeling for design, and contribute to a more flexible, automated and creative future for aerospace engineering.

\section*{Acknowledgement}
We thank Mr. Jean Fesquet and Prof. Nicolas Binder of ISAE-SUPAERO for their valuable discussions and for sharing data.
Zhen Wei and Pascal Fua are supported in part by the Swiss National Science Foundation (SNSF 200020\_204231). Zhen Wei is also supported by the French “Programme d’Investissements d’avenir” EUR-TSAE. 
Michaël Bauerheim is supported by the French Direction Générale de l’Armement through the Agence de l’Innovation de Défense in the project DECAP (Design et Contrôle par Apprentissage Profond) project.

\textbf{Disclosure of generative AI use:} Generative AI tools were used only to improve the English writing and organization of this manuscript. All technical content (methods, equations, analyses, results and references) was made and verified by the authors. No AI system is listed as an author or cited as a source. The authors reviewed all AI-assisted text and accept full responsibility for the content.

\bibliography{string1,cfd,optim,learning,vision}

\end{document}